\documentclass[lettersize,journal,twocolumn]{IEEEtran}
\IEEEoverridecommandlockouts
\usepackage{cite}
\usepackage{amsmath,amssymb,amsfonts}
\usepackage{graphicx}
\usepackage{subfloat}
\usepackage{subcaption}
\usepackage{multirow} 
\usepackage{makecell}
\usepackage{textcomp}
\usepackage{algorithm}
\usepackage{algorithmicx}
\usepackage{algpseudocode} 
\usepackage[T1]{fontenc}
\usepackage{xcolor}
\usepackage{amssymb}
\usepackage{booktabs}
\usepackage{color}
\usepackage{mathrsfs}
\usepackage{bm}
\usepackage{hyperref}

\usepackage{soul}
\sethlcolor{yellow}

\def\BibTeX{{\rm B\kern-.05em{\sc i\kern-.025em b}\kern-.08em
		T\kern-.1667em\lower.7ex\hbox{E}\kern-.125emX}}
\begin{document}
	
	\title{A Deep Learning Framework for Joint Channel Acquisition and Communication Optimization in Movable Antenna Systems}
	
	\author{Ruizhi Zhang, Yuchen Zhang,~\IEEEmembership{Member,~IEEE}, Lipeng Zhu,~\IEEEmembership{Member,~IEEE}, Ying Zhang, Rui Zhang,~\IEEEmembership{Fellow,~IEEE}
		
		\thanks{		
			Ruizhi Zhang, Ying Zhang are with School of Information and Communication Engineering, University of Electronic	Science and Technology of China, Chengdu 611731, China (e-mail: zrz\_cdut@163.com; zhying@uestc.edu.cn).
			
			Yuchen Zhang is with the Electrical and Computer Engineering Program, Computer, Electrical and Mathematical Sciences and Engineering (CEMSE), King Abdullah University of Science and Technology (KAUST), Thuwal 23955-6900, Kingdom of Saudi Arabia (e-mail: yuchen.zhang@kaust.edu.sa).
			
			Lipeng Zhu and Rui Zhang are with the Department of Electrical and Computer Engineering, National University of Singapore, Singapore 117583 (e-mail: zhulp@nus.edu.sg; elezhang@nus.edu.sg).
			
		}
		
		
	}
	
	
	
	\maketitle
	
	\begin{abstract}
		This paper presents an end-to-end deep learning framework in a movable antenna (MA)-enabled multiuser communication system. In contrast to the conventional works assuming perfect channel state information (CSI), we address the practical CSI acquisition issue through the design of pilot signals and quantized CSI feedback, and further incorporate the joint optimization of channel estimation, MA placement, and precoding design. The proposed mechanism enables the system to learn an optimized transmission strategy from imperfect channel data, overcoming the limitations of conventional methods that conduct channel estimation and antenna position optimization separately. To balance the performance and overhead, we further extend the proposed framework to optimize the antenna placement based on the statistical CSI. Simulation results demonstrate that the proposed approach consistently outperforms traditional benchmarks in terms of achievable sum-rate of users, especially under limited feedback and sparse channel environments. Notably, it achieves a performance comparable to the widely-adopted gradient-based methods with perfect CSI, while maintaining significantly lower CSI feedback overhead. These results highlight the effectiveness and adaptability of learning-based MA system design for future wireless systems.
%
	\end{abstract}
	
	
	\section{Introduction}\label{sec:intro}

To meet the demands of increasing user density and surging data volumes, multiple-input multiple-output (MIMO) technologies have been extensively studied in wireless communication systems to exploit spatial degrees of freedom (DoFs) \cite{ref1,ref2,ref3,ref4,ref5}. The associated DoFs yield beamforming and spatial multiplexing gains that significantly increase data rates, while also providing spatial diversity to improve link reliability \cite{ref6,ref7,ref8}. However, conventional MIMO architectures rely on fixed-position antennas (FPAs) arranged at predetermined locations, which limits the ability to fully exploit the continuous spatial variations of wireless channels.

To overcome these limitations, the concept of movable antenna (MA), sometimes also referred to as fluid antenna (FA) \cite{bc_ma,bc_fa}, has recently emerged as a promising approach for harnessing spatial channel variations through flexible antenna movement at transceivers\cite{ref9}. Unlike FPAs, each MA element or subarray can adjust its position to capture more favorable channel conditions. By leveraging antenna-movement DoFs, MA-aided systems can be designed to achieve substantial performance enhancements, using the same or fewer antennas than conventional FPA systems. Typical performance gains include improved signal-to-interference-plus-noise ratio (SINR)\cite{ref10,ref11}, efficient interference mitigation\cite{ref12}, flexible beamforming\cite{ref13,ref14}, and enhanced spatial multiplexing\cite{ref16}.

To exploit the spatial DoF enabled by MAs, a commonly adopted approach is joint optimization of MAs' positions and precoders, which offers promising performance enhancement by fully utilizing the additional DoFs in the spatial domain\cite{bc_ma_DoF}. 
For instance, the authors in~\cite{ref17} formulated a sum-rate maximization problem, where MAs' positions and precoders are iteratively optimized using an alternating optimization framework. In\cite{ref18}, a robust design was considered under imperfect channel state information (CSI), where joint design of antennas' positions and precoders was explored, with performance degradation due to CSI errors. Similarly,~\cite{ref19} proposed an alternating optimization framework for antenna positioning and precoding based on semidefinite relaxation and greedy updates. Moreover, in multiuser networks,\cite{ref20} integrated user scheduling, antenna movement, and precoding design into a unified optimization framework to suppress multiuser interference effectively. 
The above works collectively demonstrate that MA systems significantly outperform FPAs systems by fully unleashing the DoF in antenna positioning. Nevertheless, most existing MA studies optimize MAs' positions based on instantaneous CSI, producing antenna layouts that track specific channel snapshots. In high-mobility systems with fast-fading channels, mechanical constraints forbid repositioning within a coherence interval. Beside, frequent adjustments over short intervals also incur substantial movement overhead and energy consumption. To address these challenges, a two-timescale design paradigm can be introduced: long-term MAs' positions based on statistical CSI to amortize movement costs, coupled with short-term precoder adaptation to instantaneous CSI to preserve performance. A joint MAs' positions and precoding optimization scheme for MA systems under \textit{statistical channel} consideration was proposed in\cite{ref22}. By optimizing MAs' positions over a longer time scale using CSI aggregated across that period to define a quasi-static antenna layout and updating the precoders in each short time slot, this approach balances overall system performance with the frequency of antenna position adjustments.

However, the above works all assumed perfect knowledge of multi-path channel parameters, including perfect instantaneous CSI in snapshot adaptation and perfect statistical CSI in long-term optimization, and focus exclusively on the joint optimization of MAs' positions and precoders.
In practice, MA position optimization requires complete CSI over the entire continuous transmit and/or receive regions. Conventional channel estimation methods for FPA systems (e.g., in~\cite{ref23,ref24,ref25}) cannot be directly applied, since they only estimate the channel responses for fixed antenna positions. On the other hand, estimating channel responses across all possible positions incurs prohibitive pilot overhead and processing latency. To overcome this challenge, the authors in~\cite{CS-est-letter} proposed a novel strategy for reconstructing the full CSI between the transmit and receive regions by performing channel measurements at a finite number of MA positions. Specifically, by exploiting the sparsity of channel in the angular domain, pilots are transmitted and received at a predetermined set of locations within the Tx/Rx regions. Then, compressed sensing is applied to map these limited observations onto the sparse angular, thereby reconstructing the full channel mapping with a relatively low pilot overhead. Building on this framework, a multi-element MA channel estimation method was proposed in \cite{CS-est}. However, these approaches do not involve pilot design, and usually rely on a large set of transmit/receive measurement positions to achieve accurate channel mapping.

To address the above limitations, 
deep learning (DL) approaches, particularly deep neural networks (DNNs), have been demonstrated effective to tasks such as channel estimation, MIMO precoding, and distributed source coding (DSC) \cite{DL-WC,DL-DSC}. For channel estimation, by training DNNs on channel datasets sampled from known distributions, DL-based estimators can implicitly learn spatial correlations and nonlinear channel characteristics which are analytically intractable\cite{DL-est}. Moreover, precoding design can be formulated as an unsupervised learning problem, where a DNN is trained to map imperfect CSI to an effective precoding strategy by minimizing a performance-driven loss (e.g., the negative sum-rate)\cite{DL-precoding}. This end-to-end learning framework obviates the need for analytical channel models and enables robust optimization in complex propagation environments. DNN-based algorithms can also be used in statistical channel cases, which is particularly suitable for MA systems. First, the prime goal of statistical CSI-driven MA placement is to decrease computational overhead by optimizing positions over longer time scales, which aligns with the one-shot inference capability of a trained neural network. Second, existing gradient-based methods such as~\cite{ref22} effectively extract long-term channel statistics to derive antenna layouts but rely on iterative updates. In contrast, a data-driven model can learn the intricate mapping from statistical CSI to optimal MAs' positions offline and then instantly generate high-quality MA layouts online, eliminating the need for repeated optimization.

Building on DL advances in MIMO systems, recent works have extended DL technology to MA systems. For instance, \cite{DL-MA-est} and \cite{DL-Fas-est} exploited angular-domain sparsity and spatial correlation by sampling the channel at a limited number of MAs' positions and employing a DNN to interpolate a high-resolution channel estimate over predefined Tx/Rx grids, thereby eliminating the need for explicit per-path parameter recovery. In \cite{DL-MA-precoding} and~\cite{DL-MA-precoding2}, the authors assumed perfect CSI and treat the joint optimization of MA positions and precoders as the output of either a single DNN or two separate DNNs. These networks are trained in an end-to-end manner, with system performance metrics such as sum-rate and minimum beamforming gain as the loss function. A parallel concept has been explored in FA systems, where channel estimation, port selection, and precoders are jointly addressed. A DNN was used to extrapolate channel responses from a subset of ports to all ports, and reinforcement learning is subsequently applied to the extrapolated CSI to perform joint port selection and beamforming \cite{DRL-FAS}. Nevertheless, 
the DNN for channel estimation is trained to minimize the mean squared error (MSE) between its output and the true channel labels, whereas the joint antenna-position/precoding DNN is trained solely to enhance system performance. However, this approach incurs extra overhead by requiring an individual channel estimation network to be trained separately, and residual estimation errors can propagate into the subsequent joint antenna-position and precoding optimization, leading to cumulative performance degradation. Overall, existing DNN-based works treat channel estimation, antenna position optimization and precoder design as independent modules and rely on perfect, distortion-free outputs from preceding stages while neglecting the quantization and feedback errors introduced by uplink CSI reporting.

Moreover, in practical frequency-division duplex (FDD) systems, users acquire downlink CSI by receiving pilot signals from the BS and performing local channel estimation. The estimated CSI is then quantized and fed back to the BS, unavoidably introducing quantization errors that must be accounted for.
In conventional FDD massive MIMO systems, downlink channel estimation, CSI quantization/feedback and precoding are carried out as separate steps, incurring significant overhead and error accumulation. To overcome these limitations,~\cite{core} introduces an end-to-end DL architecture that unifies downlink pilot design, CSI quantization and limited-bit feedback, and BS-side precoder design into a sequence of connected DNN modules. By jointly training all components against a system-level performance metric (e.g., sum-rate) under a fixed feedback-bit budget, the framework bypasses explicit channel estimation and directly maps received pilot signals to precoders. This harmonized design reduces pilot overhead and feedback-bit requirements while delivering performance comparable to that of perfect-CSI linear precoding.


Inspired by \cite{core}, this paper investigates an MA-aided downlink communication system. We adopt an end-to-end mechanism that jointly addresses (i) downlink pilot design, (ii) channel estimation and quantization under feedback limit to the BS, and (iii) MAs' positions optimization together with multiuser precoding at the BS. 
The main contributions of this work are summarized as follows:

\begin{itemize}
    \item In the MA-enabled downlink system, a data-driven framework is proposed in which the BS first employs a DNN-based pilot design module to generate downlink pilots. Each user then uses a local DNN encoder to compress the received pilots into feedback bits. Upon collecting this feedback, the BS invokes a second set of interconnected DNN modules to jointly optimize MAs' positions and multiuser precoding, thereby obviating explicit channel estimation. By properly training the DNNs collectively, we jointly optimize the downlink pilots and channel estimation process, the uplink channel quantization, and the MAs' positions and multiuser precoding at the BS.
	\item Two MAs' positions optimization schemes are distinguished to reflect practical constraints: (i) MAs' positions optimization based on instantaneous channels, where MA positions are adjusted in real time to maximize throughout, and (ii) MAs' positions optimization based on statistical channels, where MA positions are updated over a longer time scale to reduce movement overhead, so as to maximize the ergodic sum rate.
	
    \item A tailored DNN training method is proposed to address key MA system challenges. In particular, a DNN-based pilot generator is introduced to allocate pilot energy across predefined antenna subsets for accurate estimation. A straight-through (ST) estimator is utilized to enable the gradient propagation through the non-differentiable binary feedback layer~\cite{STE-1,STE-2}. Continuous MAs' positions optimization is rendered tractable by discretizing the region into a number of  finite grids and recasting MA placement as a multi-class classification task, with the ST estimator handling discrete selections. The optimization of MAs' positions based on statistical CSI employs a Transformer to extract long-term channel features, enabling antenna layout design over extended time scales. By training a DNN on the estimated statistical CSI, the Transformer’s capabilities can be leveraged to infer near-optimal MA layout, ensuring robust ergodic performance.
    
    \item Comprehensive simulation results are provided to validate the proposed framework. We reveal that the end-to-end joint learning framework (i) consistently closes the gap to perfect-CSI performance across a wide range of user loads and scattering conditions, (ii) delivers significant sum-rate gains over gradient-based and codebook-based benchmarks under limited-feedback constraints, and (iii) maintains robust performance with substantially reduced pilot and feedback overhead.
    %
%
\end{itemize}

The remainder of this paper is organized as follows. Section \ref{sec: problem_statement} introduces the system and channel models, based on which the optimization problem is formulated. Section \ref{sec:Proposed mechanism} proposes a deep learning approach tailored to the formulated problem, and describes the workflow for effectively training the neural network. Section \ref{Sec:stat} extends the proposed framework to the statistical channel-based design. Section \ref{sec:Implementation Details} provides the implementation details of the DNN. Simulation results are presented in Section \ref{sec:Simulation}. Finally, Section \ref{concl} concludes this paper and sheds lights on future directions.

In this paper, the real and the imaginary parts of a vector/matrix are denoted as $\mathfrak{R}(\cdot)$ and $\mathfrak{I}(\cdot)$, respectively. $(\cdot)^T$, $(\cdot)^H$, and $(\cdot)^{-1}$ are used to denote the transpose, Hermitian transpose, and inverse of a matrix, repsectively. The identity matrix is denoted by $\textbf{I}$. In addition, $\mathbb{R}^{m\times n}$ and $\mathbb{C}^{m\times n}$ denote $m\times n$ dimensional real space and $m\times n$ dimensional complex space, repsectively. $\vert|\cdot\vert|_2$ indicates the Euclidation norm of a vector. $\mathcal{CN}(\textbf{0},\textbf{R})$ represents the zero-mean circularly symmetric complex Gaussian distribution with covariance matrix $\textbf{R}$. 
$\text{Tr}(\cdot)$, $\text{log}_2(\cdot)$, and $\mathbb{E}$ denote the trace, binary logarithm and expectation operators, respectively.  

\section{System Model And Problem Formulation} \label{sec: problem_statement}  
\subsection{MA-Aided Downlink MU-MISO System}
\begin{figure}[t!]
	\centering 
	\includegraphics[width=3.39in]{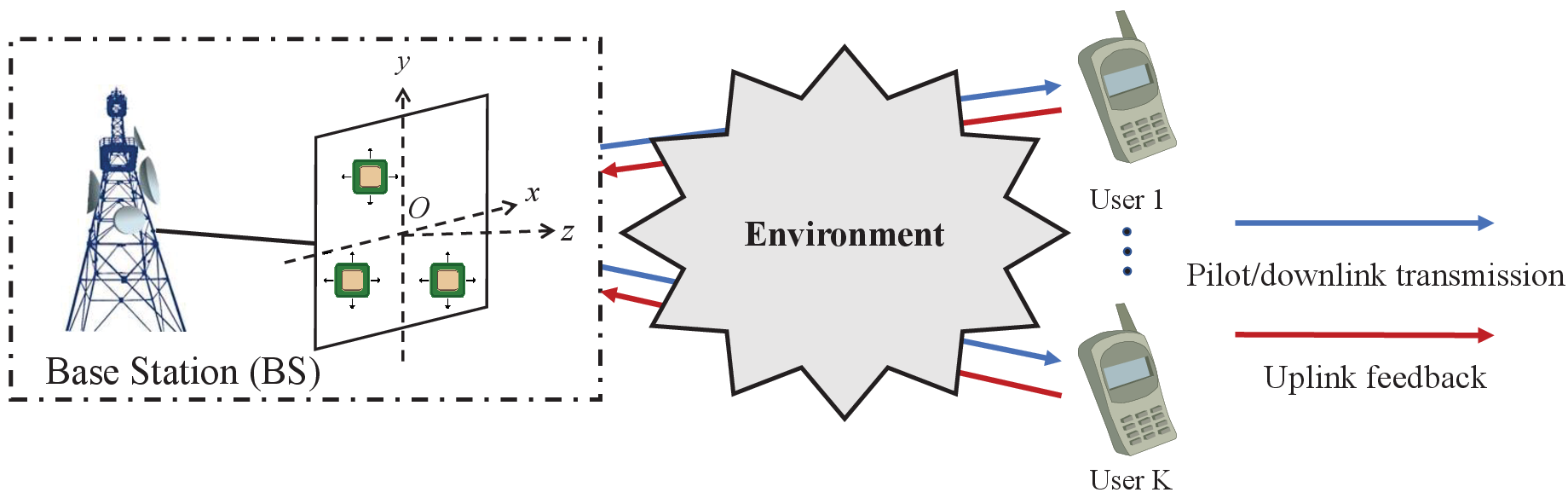} 
	\caption{The MA-aided downlink MU-MISO system.}
	\vspace{-1.5em}
	\label{fig:topo}
\end{figure}
As shown in Fig.~\ref{fig:topo}, we consider an MA-aided downlink MU-MISO system with $K$ single-antenna users. The BS is equipped with a two-dimensional (2D) array consisting of $N$ MAs, distributed over the $x$-$O$-$y$ plane.
A 3D Cartesian coordinate system is established at the BS such that the $x$-$O$-$y$ plane coincides with the antenna plane. The 2D position of the $n$-th MA in the array plane
is denoted as $\bm{a}_n = [x_n, y_n]^T \in \mathbb{R}^{2\times 1}, \forall n$, where $x_n$ and $y_n$ are the coordinates along the $x$ and $y$ axes, respectively. The antenna moving region is a rectangle area on the $x$-$O$-$y$ plane centered at the origin. Its sizes along the $x$ and $y$ axes are denoted as $S_x$ and $S_y$, respectively. 

The transmitted signal at the BS can be written as
\begin{equation}
	\textbf{x} = \sum_{k=1}^{K} \textbf{v}_ks_k = \textbf{V}\textbf{s},
\end{equation}
where $\mathbf{v}_k \in \mathbb{C}^{N \times 1}$ denotes the precoder for the $k$-th user, and $\textbf{V} = [\textbf{v}_1,...,\textbf{v}_K]\in \mathbb{C}^{N \times K}$ is the overall precoding matrix at the BS. $s_k$ represents the data stream sent to the $k$-th user, and $\mathbf{s} = [s_1, \dots, s_K]^T \in \mathbb{C}^{K\times1},$ collects all users’ data streams. Then, the received signal at the $k$-th user can be written as
\begin{equation}\label{eq:received model}
	y_k = \textbf{h}_k^H\textbf{v}_ks_k + \sum_{j\neq k} \textbf{h}_k^H\textbf{v}_js_j + z_k,
\end{equation}
where $\textbf{h}_k\in\mathbb{C}^{N\times 1}$ is the channel vector from the BS to the $k$-th user, and $z_k \sim \mathcal{CN}(0,\sigma^2)$ is the additive Gaussian noise (AWGN), with average power $\sigma^2$.

\subsection{Field-Response Based Channel Model}\label{subsec:FRV Channel}
In this subsection, a field-response-based channel model is introduced\cite{ref10}. $L_k^t$ and $L_k^r$ denote the numbers of transmit-side and receive-side channel paths between the BS to the $k$-th user, respectively. The position of $k$-th user is denoted as $\textbf{u}_k = [X_k, Y_k, Z_k]^T \in \mathbb{R}^{3\times 1}$ in its local coordinate system. The azimuth and elevation AoDs for the $l$-th transmit path are denoted as $\overline{\theta}_{k,l}^t$ and $\overline{\phi}_{k,l}^t$, respectively, while the azimuth and elevation AoAs for the $i$-th receive path are represented by $\overline{\theta}_{k,i}^r$ and $\overline{\phi}_{k,i}^r$, respectively. Then the transmit and receive field-response vectors associated with the channel from the $n$-th tranmit antenna to the $k$-th user are given by\cite{ref16}
\begin{subequations}\label{eq:FRV}
\begin{align}
    &\textbf{q}_k(\textbf{a}_n) = [
    e^{j\rho_{k,1}^t(\textbf{a}_n)},...,e^{j\rho_{k,L_k^t}^t(\textbf{a}_n)}
    ]^T \in \mathbb{C}^{L_k^t \times 1} \label{eq:FRVa}, \\
    &\textbf{f}_k(\textbf{u}_k) = [
    e^{j\rho_{k,1}^r(\textbf{u}_k)},...,e^{j\rho_{k,L_k^r}^r(\textbf{u}_k)}
    ]^T \in \mathbb{C}^{L_k^r \times 1} \label{eq:FRVb}.
\end{align}
\end{subequations}
where $\rho_{k,l}^t(\textbf{a}_n) = \textbf{a}_n^T\bm{\kappa}_{k,l}^t$ and $\rho_{k,i}^t(\textbf{u}_k) = \textbf{u}_k^T\bm{\kappa}_{k,i}^r$ denote the phase variations. Here, $\bm{\kappa}_{k,l}^t = \frac{2\pi}{\lambda}[ \cos(\overline{\theta}_{k,l}^t)\cos(\overline{\phi}_{k,l}^t), \cos(\overline{\theta}_{k,l}^t)\sin(\overline{\phi}_{k,l}^t)]^T$ and $\bm{\kappa}_{k,i}^r = \frac{2\pi}{\lambda}[    \cos(\overline{\theta}_{k,i}^r)\cos(\overline{\phi}_{k,i}^r), \cos(\overline{\theta}_{k,i}^r)\sin(\overline{\phi}_{k,i}^r), \sin(\overline{\theta}_{k,i}^r)]^T$ are the 2D transmit and 3D receive wavevectors corresponding to the $l$-th transmit path and the $i$-th receive path for the $k$-th user, respectively, where $\lambda$ is the carrier wavelength. 

Let $\bm{\Sigma}_k$ denote the path response matrix (PRM) characterizing all the transmit and receive channel paths from the BS to $k$-th user, the channel vector $\textbf{h}_k$ can be expressed as\cite{ref22}
\begin{equation}\label{eq:ins channel}
    \textbf{h}_k = \textbf{Q}_k^H \bm{\Sigma}_k \textbf{f}_k(\textbf{u}_k), 
\end{equation}
where $\textbf{Q}_k(\bm{\textbf{A}})\triangleq [\textbf{q}_k(\textbf{a}_1), ..., \textbf{q}_k(\textbf{a}_N)] \in \mathbb{C}^{L_k^t \times N}$ is the transmit field-response matrix (FRM), where $\textbf{A} = [\textbf{a}_1, \textbf{a}_2,...,\textbf{a}_N]\in\mathbb{R}^{N\times 2}$ denotes the collection of transmit antenna positions.

Equation~\eqref{eq:ins channel} makes explicit that the instantaneous channel vector \(\mathbf{h}_k\) depends on the MAs' positions \(\mathbf{A}\) through the field-response matrix \(\mathbf{Q}_k(\mathbf{A})\). Any change in MAs' positions therefore modifies \(\mathbf{Q}_k(\mathbf{A})\) and thus the observed channel. This reveals a two-way coupling: MAs' positions influence the channel estimation model used to acquire CSI, and CSI quality in turn determines the optimal MA's positions. Moreover, both CSI and MA positions jointly affect the design of downlink precoders. Consequently, pilot design, CSI estimation, MAs' positions optimization, and precoder design cannot be treated as independent tasks if this coupling is to be fully exploited. In the following, we formulate a joint optimization problem that captures the interplay between \(\mathbf{A}\), \(\mathbf{h}_k\), and the precoders.

\subsection{Problem Formulation}\label{S1:Inst_channel}

Based on the signal model in \eqref{eq:received model} and the channel model in \eqref{eq:ins channel}, the achievable rate for the \(k\)-th user is expressed as
\begin{equation}
	R_k = \log_2\left(1+\frac{\vert\textbf{h}_k^H\textbf{v}_k\vert^2}{\sum_{j\neq k}\vert\textbf{h}_k^H\textbf{v}_j\vert^2 + \sigma^2}\right).
\end{equation}
By neglecting the cost of continuous MA movement, the instantaneous $x$- and $z$-coordinates of the $N$ movable antennas, denoted by $\mathbf{a}_n$, together with the downlink precoding matrix $\mathbf{V}$, are jointly optimized to maximize the sum rate, i.e.,
\begin{equation}
    R = \sum_{k} R_k.
\end{equation}
Accurate CSI at the BS is essential to support this joint optimization. In this work, we assume that neither the BS nor the users have prior knowledge of the instantaneous channel realizations. Consequently, the CSI must be obtained through downlink training and feedback.

Specifically, the BS performs a limited number of measurements by sequentially transmitting downlink pilots from selected MAs' positions within the movable region \cite{CS-est-letter, CS-est}. Upon reception, each user applies a sparse recovery algorithm to estimate the angular domain channel parameters, which are then quantized into feedback bits and returned to the BS. According to the method in \cite{CS-est}, this limited training strategy suffices to reconstruct the channel response from any MA position to each user. The BS then uses the aggregated feedback, which implicitly encodes the estimated channel parameters, to construct the instantaneous precoding matrix and update MAs' positions. The performance of this conventional estimation and feedback scheme depends critically on two factors: the design of pilots to minimize estimation error and the feedback protocol to reduce quantization error. The detailed channel estimation procedure for the MA system is outlined as follows.


The movable region is discretized into \(M\) representative points obtained by pre-sampling~\cite{CS-est-letter, CS-est}, each point having coordinates $\{(x_m, z_m)\}$ for \(m=1,2,\dots,M\). During each downlink training round, hardware constraints allow only $N$ antennas to be positioned at different locations. Consequently, covering all \(M\) points requires $Z=\bigl\lceil\tfrac{M}{N}\bigr\rceil$
rounds of pilot transmission, indexed by \(z=1,2,\dots,Z\). In \(z\)-th round, the active antenna indices form the set \(\mathcal{M}_z\subset\{1,2,\dots,M\}\) with \(\lvert\mathcal{M}_z\rvert=N\).  The corresponding antenna positions are the points with coordinates \(\{(x_m,z_m)\mid m\in\mathcal{M}_z\}\).

For the $k$-th user, the channel vector from the \(N\) activated antennas in round \(z\) is denoted by  $\textbf{h}_{k}^{(z)}=[\,h_{k,m_{1}}^{(z)},\,h_{k,m_{2}}^{(z)},\,\dots,\,h_{k,m_{N}}^{(z)}\,]^{T}$ with entries \(h_{k,m_n}^{(z)}\) for antenna indices \(m_n\in\mathcal{M}_z\). Let $\mathbf{X}^{(z)}\in\mathbb{C}^{N\times L}$ be the pilot matrix  of length $L$ transmitted by those $N$ antennas in round $z$. Then the received pilot at user $k$ in round $z$ is $\mathbf{y}_{k}^{(z)}=(\textbf{h}_{k}^{(z)})^{H}\mathbf{X}^{(z)}+\mathbf{e}_{k}^{(z)}$, where $\mathbf{e}_k^{(z)}\in\mathbb{C}^{1\times L}$ is the noise vector for round $z$. We further define the aggregate channel vector for user $k$ by $\textbf{h}_{k}=[\,(\textbf{h}_{k}^{(1)})^{H},\,(\textbf{h}_{k}^{(2)})^{H},\,\dots,\,(\textbf{h}_{k}^{(Z)})^{H}\,]^{H}\in\mathbb{C}^{(N\,Z)\times1}$.  Likewise, stack the received signals and noise vectors over all $Z$ rounds as $\textbf{y}_{k}=[\,(\mathbf{y}_{k}^{(1)})^{H},\,(\mathbf{y}_{k}^{(2)})^{H},\,\dots,\,(\mathbf{y}_{k}^{(Z)})^{H}\,]^{T}$ and $\textbf{e}_{k}=[\,(\mathbf{e}_{k}^{(1)})^{H},\,(\mathbf{e}_{k}^{(2)})^{H},\,\dots,\,(\mathbf{e}_{k}^{(Z)})^{H}\,]^{H}$.
Then the combined received signal over $Z$ rounds can be written compactly as
\begin{equation}
    \textbf{y}_{k}=\textbf{h}_{k}^{H}
    \textbf{X}
    +
    \textbf{e}_k,
    \label{eq:combined‐received}
\end{equation}
where \(\textbf{y}_k\in\mathbb{C}^{ZL\times1}\) denotes the aggregated received pilot vector, \(\textbf{h}_k\in\mathbb{C}^{NZ\times1}\) denotes the stacked channel vector, and \(\textbf{e}_k\in\mathbb{C}^{ZL\times1}\) denotes the combined noise vector. The pilot matrix \(\textbf{X}\) is block diagonal and can be written as  
\begin{equation}\label{diagX}
    \textbf{X}
    =
    \mathrm{blkdiag}\bigl(
        \mathbf{X}^{(1)},\mathbf{X}^{(2)},\dots,\mathbf{X}^{(Z)}
    \bigr)
    \in\mathbb{C}^{(NZ) \times (ZL)}.
\end{equation}



After \(Z\) downlink training rounds, each user jointly processes the collected observations to recover the sparse channel parameters and then transmits this information to the base station using a total of \(B\) feedback bits
\begin{equation}
	\textbf{q}_k = \mathcal{F}_k(\textbf{y}_k),
\end{equation}
where the function $\mathcal{F}_k:\mathbb{C}^{1\times (ZL)}\rightarrow {\pm 1}^B$ represents the feedback scheme adopted at user $k$.

Finally, the BS aggregates the feedback bits from all $K$ users into $\textbf{Q} \triangleq [\textbf{q}_1,\textbf{q}_2,...,\textbf{q}_K]$, and performs two operations in sequence:
\begin{itemize}
    \item MAs' positions optimization
\begin{equation}
\begin{aligned}
    \mathbf{A} =& f_{\mathrm{AP}}(\mathbf{Q})
\end{aligned}
\end{equation}
   where $f_{\mathrm{AP}}(\cdot):\{\pm 1\}^{KB}\longrightarrow\mathbb{R}^{N\times2}$ denotes the MAs' positions optimization scheme adopted at the BS.
   

   \item Precoder design
\begin{equation}
\begin{aligned}
    \mathbf{V} =& f_{\mathrm{BF}}(\mathbf{Q},\mathbf{A})
\end{aligned}
\end{equation}
where $f_{\mathrm{BF}}(\cdot):\bigl[\{\pm1\}^{\,K\,B},\mathbb{R}^{N\times2}\bigr]
\longrightarrow \mathbb{C}^{\,N\times K}$ denotes the precoding scheme. The dependence of \(f_{\mathrm{BF}}\) on \(\mathbf{A}\) reflects the fact that spatial correlations among MA elements vary with their locations.
\end{itemize}

Based on the above analysis, the instantaneous sum-rate maximization problem for the MA-enabled system can be formulated as
\begin{subequations}\label{prob:p1}
\begin{align}
(\mathcal{P}1): \quad 
& \max_{
\textbf{X},\, \{\mathcal{F}_k(\cdot)\}_{\forall k},\, f_{\text{AP}}(\cdot), f_{\text{BF}}(\cdot)} \; R \notag \\ 
\text{s.t.} \quad 
& \; \mathbf{A} = f_{\text{AP}}(\mathbf{Q}), \\
& \; \mathbf{V} = f_{\text{BF}}(\mathbf{Q},\mathbf{A}), \\
& \; \textbf{q}_k = \mathcal{F}_k(\textbf{y}_k), \forall k \in \{1,...,K\}, \\
& \; \mathrm{Tr}(\mathbf{V} \mathbf{V}^H) \leq P_{\text{max}}, \\
& \; \|\textbf{X}_l\|_2^2 \leq P_{\text{max}}, \quad \forall l \in \{1,...,ZL\}, \\
& \sqrt{(x_i - x_j)^2 + (z_i - z_j)^2} \geq \lambda/2, \forall i \neq j. \label{spacing constraint}
\end{align}
\end{subequations}
In this formulation, the downlink pilot matrix \(\tilde{\mathbf{X}}\), the user feedback schemes \(\{\mathcal{F}_k\}_{\forall k}\), the MAs' positions optimization scheme \(f_{\mathrm{AP}}\) and the BS precoding scheme \(f_{\mathrm{BF}}\) are jointly optimized to maximize the instantaneous sum rate. Problem~\(\mathcal{P}1\) couples pilot design, CSI feedback, MA positioning and precoding into a holistic optimization. This joint design is challenging because it requires distributed quantization of channel information and entails a tight interdependence between MAs' positions and precoders. Simple heuristic, such as independent codebook-based quantization of the channel vector at each user, is likely to be far from the optimum. To address these challenges, this paper introduces an efficient solution based on data-driven DL and modular neural network architectures, enabling structured end-to-end optimization of all functional components.

\section{Joint Design Using Deep Learning}\label{sec:Proposed mechanism}

In this section, we elaborate on the DNN-based modeling of the MA-enabled downlink system introduced in Section \ref{sec: problem_statement}. The overall end-to-end process comprises three sequential phases: (i) downlink channel training and uplink CSI feedback, (ii) MA's positions optimization based on the quantized CSI feedback, and (iii) precoding design under the optimized fixed MAs' positions. Furthermore, we then detail the training methodology for the proposed modular network architecture, which jointly optimizes the downlink training pilots \(\textbf{X}\), the feedback scheme \(\mathcal{F}_k(\cdot)\) for each user, the MA positioning scheme \(f_{\mathrm{AP}}(\cdot)\), and the BS precoding scheme \(f_{\mathrm{BF}}(\cdot)\).

\subsection{DNN Representation}\label{subsec: DNN Rep}

To accurately model the MA-enabled multi-user downlink communication system introduced in section \ref{sec: problem_statement}, we need to model each component as a neural network, which contains downlink training and user feedback scheme for channel estimation, MAs' positions optimization and precoder design with instantaneous CSI.

\textit{1) Downlink Pilot Training of MA:} We begin by modeling the first part of the downlink training phase, in which the BS sends training pilots $\textbf{X}\in\mathbb{C}^{(NZ)\times (ZL)}$ in $L \times Z$ downlink training, and consequently the $k$-th user observes $\textbf{y}_k = \textbf{h}_k^H\textbf{X} + \textbf{e}_k$. 
By considering $\textbf{h}_k$ as the input, it is easy to see that the received signal $\textbf{y}_k$ at each user in the total downlink training phase can be modeled as the output of a fully-connected neural network layer with the linear activation function, in which the weight matrix is $\textbf{X}$ and the bias vector is zero, followed by an additive zero-mean noise with variance $\sigma^2$. In this formulation, the neural network is structured as a single-layer linear model with $\textbf{X}$ as its weight matrix, and zero bias, aligning closely with the physical representation of pilot transmissions and additive zero-mean noise with variance $\sigma^2$.

Due to the block-diagonal structure of the pilot matrix $\tilde{\mathbf{X}}$ as defined in \eqref{diagX}, localized training can be performed within each sub-array $\tilde{\mathbf{X}}^{(z)}$, which facilitates parallel signal processing and hardware implementation. To ensure that the learned matrix adheres to this structural constraint, a post-processing step is applied to project it onto the block-diagonal space. The specific implementation details of this projection will be elaborated in Section~\ref{sec:Implementation Details}.

To comply with power constraints on each downlink training, a weight constraint is enforced under which each column of $\textbf{X}$ satisfies $\vert|\textbf{X}_l\vert|^2_2 \leq P_{\rm max}, \forall l \in \{1,...,ZL\}$. This ensures that the instantaneous transmission power in each downling transmission does not exceed the allowed maximum power $P_{\text{max}}$. To ensure this power constraint, we always normalize the updated $\textbf{X}$ in the training process such that $\vert|\textbf{X}_l\vert|_2^2 = P_{\rm max}$.

%


\textit{2) Uplink Feedback:}
Upon completion of the downlink training phase, user \(k\) obtains the aggregated pilot observation \(\widetilde{\mathbf{y}}_k\). The task of each user is to extract and quantize the relevant channel features into a \(B\)-bit feedback vector.
We model the feedback generation process at user $k$ as a feature extraction and quantization task, which can be implemented using an $R$-layer neural network (e.g., a standard feedforward network or a convolutional neural network) at each user side
\begin{equation}
    \textbf{q}_k = \text{sgn}\left(\textbf{W}_R^{(k)}\sigma_{R-1}\left(\cdots\sigma_1\left(\textbf{W}_1^{(k)}\overline{\mathbf{y}}_k + \textbf{b}_1^{(k)}\right)\cdots\right) + \textbf{b}_R^{(k)}\right),
\end{equation}
where $\textbf{q}_k \in \{\pm 1\}^B$, $\{\textbf{W}_r^{(k)}, \textbf{b}_r^{(k)}\}_{r=1}^R$ are the set of the trainable parameters for user $k$, $\sigma_r$ is the activation function for the $r$-th layer, and the sign function $\text{sgn}(\cdot)$ is the activation function of the last layer to generate binary feedback bits for each elements of $\textbf{q}_k$. At the hidden layers, we adopt the rectified linear unit (ReLU) activation function, i.e., $\sigma_r(\cdot) =\max(\cdot,0)$. As defined in~\eqref{yk}, $\overline{\mathbf{y}}_k$ represents the concatenation of real and imaginary parts of the received pilot matrix $\textbf{y}$ as input features to the network
\begin{equation}\label{yk}
    \overline{\mathbf{y}}_k = \left[\mathfrak{R}(\textbf{y}_k)^T, \mathfrak{I}(\textbf{y}_k)^T\right]^T.
\end{equation}


The feedback vector $\mathbf{q}_k$ encapsulates a compact representation of the effective downlink channel observed by user $k$, specifically designed to support downstream processing at the BS, which the BS leverages for downstream tasks including MA positioning and precoder design.

\textit{3) MA Positioning:}
Assuming an error‑free uplink feedback link, the BS aggregates the feedback bits from all users to determine the optimal MAs' positions. The resulting antenna configuration then serves as critical input to the downlink precoder design.

In an MA-aided communication system, the objective is to dynamically select the optimal $N$ antenna positions from a continuous feasible region $\mathcal{R} \subset \mathbb{R}^2$. Directly optimization over this continuous domain results in infinite-dimensional problem.
To address this, \(\mathcal{R}\) is uniformly discretized into a two‑dimensiona grid consisting of \(G = G_x \times G_z\) candidate points, where \(G_x\) and \(G_z\) denote the number of grid locations along the horizontal and vertical axes, respectively. The discrete feasible set of antenna positions is then defined as  
\begin{equation}
\mathcal{G} = \left\{ (x, z) \;\middle|\;
\begin{aligned}
&x = 0,\, d,\, \ldots,\; (G_x - 1)d, \\
&z = 0,\, d,\, \ldots,\; (G_z - 1)d
\end{aligned}
\right\}.
\end{equation}
where $d$ denotes the spatial resolution (i.e., the grid spacing). 


To choose \(N\) out of these \(G\) points using feedback, we introduce a binary selection mask \(\mathbf{m}\in\{0,1\}^G\) with exactly \(N\) ones, which is generated by a \(U\)-layer neural network
\begin{equation}
\textbf{m} = \overline\sigma_U\Bigl(\overline{\mathbf{W}}_U,\sigma_{U-1}\bigl(\cdots\sigma_1(\overline{\mathbf{W}}_1\mathbf{Q} + \overline{\mathbf{b}}_1)\cdots\bigr) + \overline{\mathbf{b}}_U\Bigr),
\end{equation}
where $\overline{\mathbf{W}}_u$ and $\overline{\mathbf{b}}_u$ are the weight matrix and bias vector of the $u$-th layer, respectively, and $\overline\sigma_u(\cdot)$ is the non-linear activation function. To ensure exactly $N$ active selections, and \(\bar\sigma_U(\cdot)\) is the final activation function designed to produce a sparse binary output with exactly \(N\) active entries.  




To enable gradient‑based optimization of discrete antenna selection, we first compute a soft selection probability vector \(\mathbf{p}_{\mathcal{G}}\in\mathbb{R}^G\) from the pre‑activation vector \(\boldsymbol{\alpha}=(\alpha_1,\dots,\alpha_G)^T\) using a temperature-controlled Boltzmann distribution \cite{soft}
\begin{equation}\label{T-softmax}
p_g = \frac{\exp\left( \alpha_g / \tau \right)}{\sum_{g'=1}^{G} \exp\left( \alpha_{g'} / \tau \right)}, \quad \forall g \in {1, \dots, G},
\end{equation}
where $\tau > 0$ is a temperature parameter that regulates the concentration of the antenna selection probability distribution $\mathbf{p}_{\mathcal{G}} = [p_1, \dots, p_G]^T$. A hard binary mask \(\mathbf{m}\in\{0,1\}^G\) with exactly \(N\) ones is then obtained by selecting the indices of the \(N\) largest entries in \(\mathbf{p}_{\mathcal{G}}\):
\begin{equation}
m_g =
\begin{cases}
1, & \text{if } g \in \mathcal{I}_N\big( \mathbf{p}_{\mathcal{G}} \big), \\
0, & \text{otherwise},
\end{cases}
\end{equation}
where $\mathcal{I}_N(\cdot)$ returns the set of indices corresponding to the top‑\(N\) values.

Accordingly, the final-layer activation function $\overline{\sigma}_U(\cdot)$ is implemented as a composition of a temperature‑scaled softmax and a top-\(N\) selection operator
\begin{equation}
\overline{\sigma}_U(\boldsymbol{\alpha}) = \mathcal{H}_N \left( \left[ \frac{\exp(\alpha_g / \tau)}{\sum_{g'=1}^{G} \exp(\alpha_{g'} / \tau)} \right]_{g=1}^{G} \right),
\end{equation}
Here, $\mathcal{H}_N(\cdot)$ produces a binary mask with ones at the indices of the \(N\) largest entries.
For hidden layers of the selection network, we also employ the ReLU activation function. 
Given the resulting mask \(\mathbf{m}\), we can obtain optimized MAs' positions as $\textbf{A} = \{ \mathbf{a}_i \in \mathcal{G} \mid m_i = 1 \}$.

\textit{4) Precoding Design:}
In the precoding phase, after error‑free uplink feedback and MA positioning optimization, the BS constructs the downlink precoding matrix using both the optimized antenna layout and the aggregated quantized feedback from all \(K\) users. Each user \(k\) returns a \(B\)-bit feedback vector \(\mathbf{q}_k\), which are concatenated into $\textbf{Q}$. This feedback implicitly captures each user’s effective downlink channel under the selected MA configuration.  

To map this compact feedback and antenna configuration into a high-quality downlink precoding matrix, a DNN-based precoding module is proposed to learn an approximation of the optimal mapping:
\begin{equation}
f_{\text{BF}}:\left\{\textbf{Q}, \mathbf{A}\right\}\rightarrow\mathbb{C}^{N\times K},
\end{equation}
Similar with the user-side, the precoding design module is implemented as a \(T\)-layer neural network. The downlink precoding matrix \(\mathbf{V}\in\mathbb{C}^{N\times K}\) is vectorized as
\begin{equation}
    \textbf{v} = \left[\text{vec}(\mathfrak{R}(\textbf{V}))^T, \text{vec}(\mathfrak{I}(\textbf{V}))^T\right]^T,
\end{equation}
can be written as:
\begin{equation}
    \textbf{v} = \tilde{\sigma}_T\left(\tilde{\textbf{W}}_T\tilde{\sigma}_{T-1}\left(\cdots\tilde{\sigma}_1\left(\tilde{\textbf{W}}_1\mathcal{J} + \tilde{\textbf{b}}_1\right)\cdots\right) + \tilde{\textbf{b}}_T\right),
\end{equation}
where for each layer \(t=1,\dots,T\), \(\tilde{\mathbf{W}}_t\) and \(\tilde{\mathbf{b}}_t\) are the trainable weight matrix and bias vector, and \(\tilde\sigma_t(\cdot)\) denotes the activation function.  The network input \(\mathcal{J}\) concatenates the aggregated feedback \(\mathbf{Q}\) and the optimized MA positions \(\mathbf{A}\): $\mathcal{J} = \left[\text{vec}(\textbf{Q})^T, \text{vec}(\textbf{A})^T\right]^T$.
Finally, to satisfy the transmit power constraint, a normalized layer with activation function
\begin{equation}
\tilde{\sigma}_T(\cdot) = \sqrt{P_{\text{max}}}\;\frac{(\cdot)}{\|\cdot\|_2}.
\end{equation}
is employed at the last layer of the neural network, while all hidden layers employ the ReLU activation function.

\begin{figure*}[!t]
  \vspace{-0.2em}
  \centering
  \includegraphics[width=0.78\textwidth]{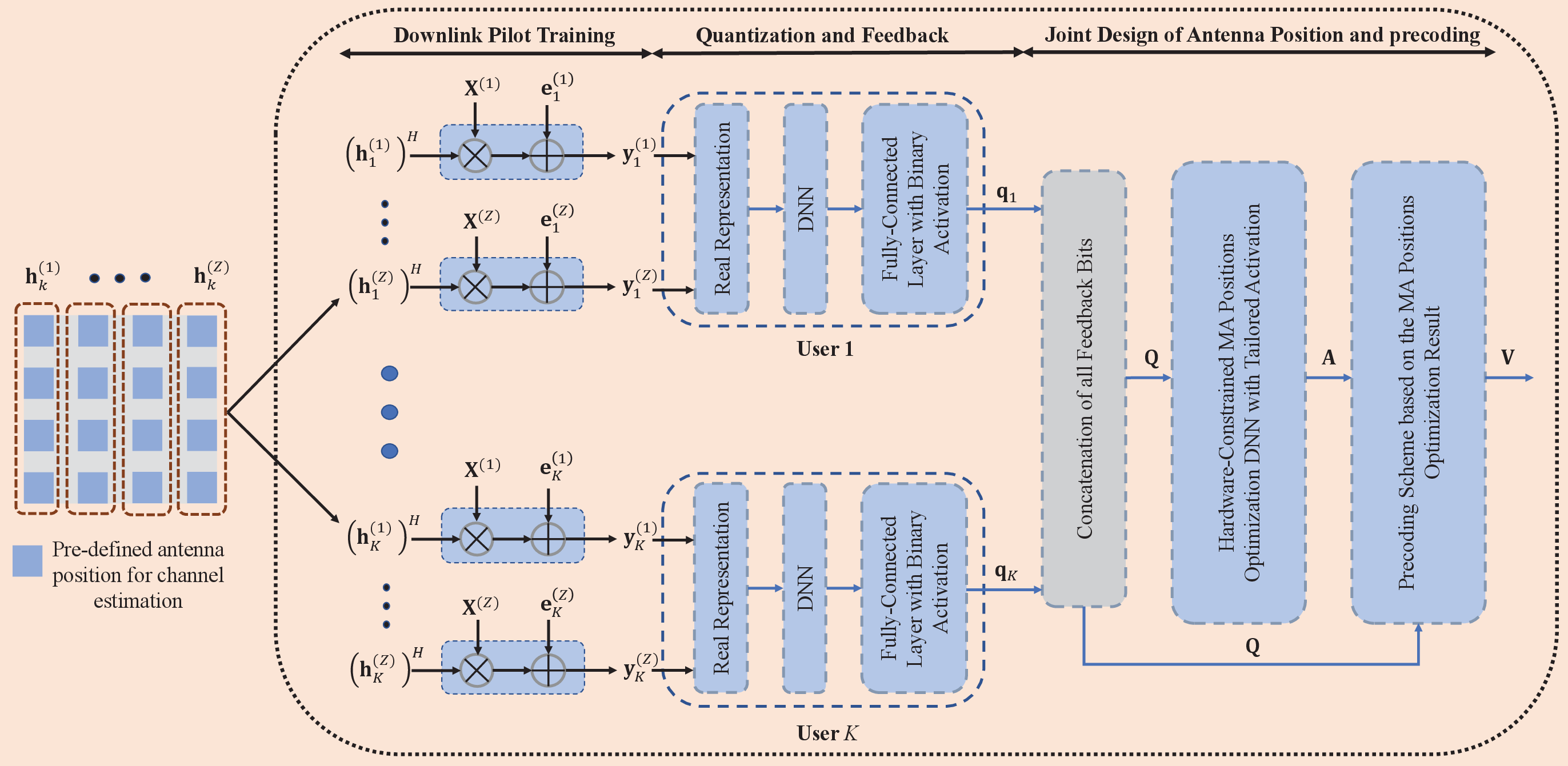}
  \caption{Illustration of the proposed end-to-end DL framework for a MA-enabled downlink communication System.}
  \label{fig:system}
  \vspace{-1.5em}
\end{figure*}

The block diagram of the overall proposed neural network architecture that represents an end-to-end MA-aided system is illustrated in Fig. \ref{fig:system}. In this neural network the trainable parameters are the neural network parameters $\Theta_R^{(k)}\triangleq \{\textbf{W}_r^{(k)}, \textbf{b}_r^{(k)} \}_{r=1}^R$ at the user side, and the training pilot matrix $\textbf{X}$, the neural network parameters $\Theta_A\triangleq \{\overline{\textbf{W}}_u, \overline{\textbf{b}}_u\}_{u=1}^U$, $\Theta_T\triangleq \{\tilde{\textbf{W}}_t, \tilde{\textbf{b}}_t\}_{t=1}^T$ at the BS side.

\subsection{DNN Training Process}
We now describe the training of the DNN architecture in Fig. 2 for the sum rate maximization objective as stated in the following
\begin{equation}\label{eq:loss function}
\max_{\textbf{X},\Theta_R^{(k)}, \Theta_A, \Theta_T} \mathbb{E}_{\mathbf{H}} \left[ \sum_{k} \log_2\left(1+\frac{\vert\textbf{h}_k^H\textbf{w}_k\vert^2}{\sum_{j\neq k}\vert\textbf{h}_k^H\textbf{w}_j\vert^2 + \sigma^2}\right)\right]
\end{equation}
where the expectation is over the distribution of the channels, i.e., $\textbf{H} \triangleq \left[\textbf{h}_1,...,\textbf{h}_K\right]^H$. The parameter space consists of the training pilot matrix, the users' feedback schemes, the MA position optimization scheme and the precoding scheme.

To guide the training of the overall neural architecture, we define the loss function based on the system's instantaneous performance. Specifically, for each realization of the instantaneous channel in a mini-batch, we compute the achievable sum rate $R$ based on the current antenna position, precoding matrix, and feedback bits. The training loss is then defined as the negative average sum rate across the batch:
\begin{equation}
\mathcal{L} = -\frac{1}{|\mathcal{B}|} \sum_{b \in \mathcal{B}} \sum_{k=1}^K R_k^{(b)}
\end{equation}
where $|\mathcal{B}|$ is the batch size and $R_k^{(b)}$ is the rate of user $k$ in the $b$-th sample. This loss directly encourages the network to learn representations that maximize system throughput. While the above design enables flexible antenna selection, it does not consider the \textit{minimum distance constraint} between selected antennas, which is essential to avoid strong mutual coupling. To address this, we introduce a regularization term that penalizes selections violating the minimum spacing $\lambda/2$. We define the penalty function for enforcing the minimum inter-antenna spacing as
\begin{equation}
\mathcal{L}_{\text{dist}} = -\frac{1}{|\mathcal{B}|}\sum_{b \in \mathcal{B}}\sum_{i < j} \max\left(0,\, \lambda/2 - \|\mathbf{r}_i^{(b)} - \mathbf{r}_j^{(b)}\|_2 \right)^2.
\end{equation}
where $\textbf{r}_i^{(b)}, \textbf{r}_j^{(b)}$ denote the $i$-th and $j$-th positions, respectively, from the MA position optimization result of the $b$-th sample. Then the final training loss becomes
\begin{equation}
\mathcal{L}_{\text{total}} = \mathcal{L} + \beta \cdot \mathcal{L}_{\text{dist}},
\end{equation}
where $\beta$ is a parameter that balances system performance and spatial feasibility.

In the feedback stage, each user quantizes the received pilot response into a $B$-bit sequence composed of $\pm 1$ values. Due to the fact that the derivative of the output of a binary thresholding neuron is almost everywhere, the conventional back-propagation method cannot be directly used to train the neural layers prior to that binary layer. To enable gradient-based learning while preserving the discrete nature of the feedback, we adopt an ST estimator strategy\cite{STE-1,STE-2}. In particular, the sign function $\text{sgn}(u)$ in the back-propagation phase can be replaced approximately by a sigmoid-adjust STE with slope annealing estimator
\begin{equation}
    2\text{sigm}(\omega^{(i)}u) - 1 = \frac{2}{1+\exp(-\omega^{(i)}u)}-1
\end{equation}
where $\text{sigm}(\omega^{(i)}u)$ is the sigmoid function, and $\omega^{(i)}$ is the annealing factor in the $i$-th epoch satisfying $\omega^{(i)} \geq \omega^{(i-1)}$ to gradually sharpen the quantization. This trick ensures the forward process uses valid binary outputs while maintaining a non-zero gradient for backpropagation.

In the antenna position optimization module, a similar ST-based strategy is adopted
\begin{equation}
\mathbf{m} = \mathbf{p}_{\mathcal{G}} + \mathcal{D}\left( \mathbf{m} - \mathbf{p}_{\mathcal{G}} \right),
\end{equation}
where $\mathcal{D}(\cdot)$ represents the stop-gradient operator, which blocks gradient flow through its input. This formulation ensures that $\mathbf{m}$ functions as a hard selection mask during the forward pass, while preserving differentiability during backpropagation.
\begin{algorithm}[!t]
\caption{DNN-Based MA-Aided Communication Systems}
\label{alg:joint_ma_dnn}
\begin{algorithmic}[0]
\item[\underline{\textbf{Input:}}]  Channel distribution $\mathcal{H}$, candidate grid $\mathcal{G}$ ($|\mathcal{G}|=G$), grid spacing $d$, number of MAs $N$, number of users $K$, number of feedback bits $B$, pilot slots $(L,Z)$, temperature $\tau$, power limit $P_{\max}$, training dataset size $\mathcal{N}_t$, validation dataset size $\mathcal{N}_v$, batch size $|\mathcal{B}|$, number of total epochs $N_{\rm ep,max}$, annealing rate $\omega^{(0)}$, learning rate $\eta^{(0)}$.
\item[\underline{\textbf{Output:}}] Learned parameters $\;\tilde{\mathbf{X}},\{\Theta_R^{(k)}\}_{k=1}^K,\Theta_A,\Theta_T$.

\item[\underline{\textbf{Initialize:}}]
\State \quad $i \leftarrow 0$;
\State \quad Generate training set $\mathcal{S}_t$ of size $\mathcal{N}_t$ and validation set $\mathcal{S}_v$ of size $\mathcal{N}_v$ based on $\mathcal{H}$;
\State \quad Randomly initialize $\tilde{\mathbf{X}}$, $\{\Theta_R^{(k)}\}$, $\Theta_A$, $\Theta_T$;
\State \quad best\_rate $\leftarrow$ average sum rate on $\mathcal{S}_v$;

\item[\underline{\textbf{DNN Training:}}]
\item[\textbf{while} $i < N_{\rm ep,max}$ \textbf{do}]
    \For{each mini-batch $\{\tilde{\mathbf{h}}_k^{(b)}, \mathbf{H}^{(b)}\}_{b\in\mathcal{B}} \in \mathcal{S}_t$}

        \Statex \quad \underline{\textbf{Phase I: Downlink Training}}
        \Statex \quad\quad Transmit pilots $\tilde{\mathbf{X}}$ over $L \times Z$ downlink training;
        \Statex \quad\quad Each user $k$ receives $\tilde{\mathbf{Y}}_k^{(b)} = (\tilde{\mathbf{h}}_k^{(b)})^H \tilde{\mathbf{X}}^{(b)} + \tilde{\mathbf{e}}_k^{(b)}$;

        \Statex \quad \underline{\textbf{Phase II: Uplink Feedback}}
        \Statex \quad\quad Form feature $\overline{\mathbf{y}}_k^{(b)} = [\Re(\tilde{\mathbf{Y}}_k^{(b)})^T, \Im(\tilde{\mathbf{Y}}_k^{(b)})^T]^T$;
        \Statex \quad\quad $\mathbf{q}^{(b)}_k = \mathcal{F}_k(\overline{\mathbf{y}}^{(b)}_k; \Theta_R^{(k)}) \in \{\pm1\}^B$ (STE);

        \Statex \quad \underline{\textbf{Phase III: MA Position Optimization}}
        \Statex \quad\quad Stack $\mathbf{Q}^{(b)} = [(\mathbf{q}_1^{(b)})^T; \dots; (\mathbf{q}_K^{(b)})^T]^T$;
        \Statex \quad\quad Compute logits $\boldsymbol\alpha^{(b)} = \mathrm{FC}_A(\mathbf{Q}^{(b)}; \Theta_A)$;
        \Statex \quad\quad $\mathbf{p}^{(b)}_{\mathrm{soft}} = \mathrm{softmax}(\boldsymbol\alpha^{(b)} / \tau)$;
        \Statex \quad\quad $\mathbf{m}^{(b)} = \mathrm{GreedySelect}(\mathbf{p}^{(b)}_{\mathcal{G}}, N)$;
        \Statex \quad\quad Hard-mask via STE: $\tilde{\mathbf{m}}^{(b)} = \mathbf{p}^{(b)}_{\mathcal{G}} + \mathcal{D}(\mathbf{m}^{(b)} - \mathbf{p}^{(b)}_{\mathcal{G}})$;

        \Statex \quad \underline{\textbf{Phase IV: Downlink Precoding}}
        \Statex \quad\quad Fuse feedback \& mask: $\mathcal{J}^{(b)} = [(\mathbf{Q}^{(b)})^T; (\mathbf{m}^{(b)})^T]^T$;
        \Statex \quad\quad Get precoding output: $\textbf{v}^{(b)} = f_{BF}(\mathcal{J}^{(b)}; \Theta_T)$;
        \Statex \quad\quad Apply mask and normalize:
        \Statex \quad\quad\quad $\textbf{v}^{(b)} \leftarrow \textbf{v}^{(b)} \odot \mathbf{m}$;
        \Statex \quad\quad\quad $\textbf{v}^{(b)} \leftarrow \sqrt{P_{\max}}\, \textbf{v}^{(b)} / \|\textbf{v}^{(b)}\|_2$;

        \Statex \quad \underline{\textbf{Loss and Update}}
        \Statex \quad\quad Compute per-user rates $R_k^{(b)}$;
        \Statex \quad\quad $\mathcal{L} = -\frac{1}{|\mathcal{B}|} \sum_{b,k} R_k^{(b)}$;
        \Statex \quad\quad Back-propagate and update all parameters;

        \Statex \quad \underline{\textbf{Validation}}
        \Statex \quad\quad current\_rate $\leftarrow$ average sum rate on $\mathcal{S}_v$;
        \Statex \quad\quad \textbf{if} current\_rate $>$ best\_rate
            \Statex \quad\quad\quad  Save all parameters;
            \Statex \quad\quad\quad  best\_rate $\leftarrow$ current\_rate;
        \Statex \quad\quad\textbf{end if}
        \Statex \quad\quad $i \leftarrow i + 1$;
        \Statex \quad \textbf{Increase} annealing rate;
        \Statex \quad \textbf{Decrease} learning rate;
    \EndFor
\item[\textbf{end while}] 
\vspace{-0.3em}
\end{algorithmic}
\end{algorithm}

\section{Two Time-scale Design Framework}\label{Sec:stat}

In practical scenarios, frequent antenna movement often incurs additional overhead, and the channel coherence time may be insufficient for antenna movement, particularly for mechanically driven MAs operating under fast fading conditions. To address this issue, we consider extending the proposed design to leverage statistical CSI, which allows MAs' positions to be optimized over longer timescales, thereby significantly reducing movement energy comsumption and overhead.

\subsection{Statistical Channel Model}

Based on the instantaneous channel model introduced in Section~\ref{subsec:FRV Channel}, a statistical channel model can be developed. The BS position is fixed long-term at high altitude, surrounded mainly by large dominant scatterers that chiefly determine the AoDs of NLoS paths. Each user may move within a local region. Since the movement ranges of the BS antenna and users are generally much smaller than the signal propagation distances between BS/user and their dominant scatterers, and the scatterers are assumed fixed, the AoDs remain essentially constant. Under the far-field assumption, the AoAs at the user side also vary minimally despite user movement, allowing to approximate them as unchanged for each path. Therefore, the transmit FRM $\mathbf{Q}_k$ and PRM $\bm{\Sigma}_k$ can be regarded as approximately constant. However, the receive FRV $\mathbf{f}_k(\mathbf{u}_k)$ may change rapidly, where the phase shifts $\rho_{k,i}^r(\mathbf{u}_k), \forall k,i$, are modeled as independent and identically distributed (i.i.d.) random variables uniformly distributed over $[0, 2\pi)$. Following \cite{ref22}, the statistical channel model for $\mathbf{h}_k$ is given by
\begin{equation}\label{eq:stat model}
    \textbf{h}_k = \textbf{Q}_k^H\bm{\psi}_k, \bm{\psi}_k\sim \mathcal{CN}(\bm{0}_{L_k^t\times 1}, \text{Diag}(\textbf{b}_k)), \forall k.
\end{equation}
where $\bm{\psi}_k = \bm{\Sigma}_k \textbf{f}_k(\textbf{u}_k) \in \mathbb{C}^{L_k^t\times 1}$ denotes the transmit path-response vector for user $k$. Hence, we can obtain that $\psi_{kl} = \sum_{i=1}^{L_k^r} \Sigma_{k,li} \exp(j\rho_{k,i}^r(\textbf{u}_k))$. Since $L_k^r$ is generally large in practice due to the rich scattering enviroment around user, $\psi_{kl}$ can be approximately modeled as a CSCG random variable \cite{ref22}. Specially, we have $\psi_{kl} \sim \mathcal{CN}(0, b_{kl})$, where $b_{kl}$ denotes the expected path-response power of $l$-th transmit channel path for user $k$. Moreover, given that $L_k^r$ is large and phases of $\Sigma_{k,li}, \forall l,i$ are independent and uniformly distributed within $(0, 2\pi]$, we can approximate $\mathbb{E}[\psi_{kl}^* \psi_{kl}] = 0$. Hence, $\textbf{b}_k \in [b_{k1},...,b_{k L_k^t}]^T \in \mathbb{R}^{L_k^t \times 1}$ is defined as the transmit path-response power vector for user $k$, which can be regarded as the angular power spectrum for the channel of user $k$ with the BS and characterizes the average power distribution on the multi-path channel in the angular domain.

\subsection{Ergodic Sum Rate Maximization under Statistical Channel}

To alleviate the stringent requirement of reconfiguring MAs' positions in each short channel coherence time, we further consider maximizing the ergodic sum rate via a two-timescale design approach. Specifically, precoding matrix $\textbf{V}$ is optimized based on the instantaneous CSI to maximize the instantaneous sum rate, while antenna positions are designed over a relatively longer period to improve the ergodic sum rate.
In this way, the downlink pilot training, user feedback, and precoding which is same with the instantaneous CSI scenario are still required, as described in Section~\ref{S1:Inst_channel}.
In contrast, the BS optimizes the antenna position based on the feedback bits over a period of time which implicitly contains the statistical CSI information, with the function given by
\begin{equation}
	\mathbf{A} = f^S_{\text{AP}}\bigl(\mathbf{Q}^{\text{all}}\bigr), \quad
	f^S_{\text{AP}}:\{\pm 1\}^{T_{\delta} K B} 
	\;\longrightarrow\; \mathbb{R}^{N\times 2}
\end{equation}
where $T_{\delta}$ denotes the number of feedback intervals (or time slots) over which the BS collects feedback bits to capture the statistical CSI. Here, $f^S_{\text{AP}}$ refers to the MAs' positions optimization scheme based on statistical CSI, and $\mathbf{Q}^{\text{all}}$ represents the aggregate feedback bits collected during this period.

Similar to~\eqref{prob:p1}, the problem of maximizing the ergodic sum rate of a MA-enabled communication system can be summarized as
\begin{subequations}\label{prob:p2}
\begin{align}
(\mathcal{P}2): \quad 
& \max_{f^S_{\text{AP}}(\cdot)} \; \mathbb{E}_{\mathbf{H}} \left[ \max_{\textbf{X},\, \{\mathcal{F}_k(\cdot)\}_{\forall k},\, f_{\text{BF}}(\cdot)} \; R \right] \notag \\ 
\text{s.t.} \quad 
& \; \textbf{A} = f^S_{\text{AP}}(\textbf{Q}^{\text{all}}), \\
& \; \mathbf{V} = f_{\text{BF}}(\mathbf{Q},\mathbf{A}), \\
& \; \mathbf{q}_k = \mathcal{F}_k\left(\{\textbf{y}_k(z) \}_{z=1}^{Z} \right), \forall k, \\
& \; \mathrm{Tr}(\mathbf{V} \mathbf{V}^H) \leq P_{\text{max}}, \\
& \; \|\textbf{X}_l\|_2^2 \leq P_{\text{max}}, \quad \forall l, \\
& \;\sqrt{(x_i - x_j)^2 + (z_i - z_j)^2} \geq \lambda/2, \forall i \neq j.
\end{align}
\end{subequations}
The MAs' positions should be optimized based on historical data under a given channel distribution, aiming to obtain a solution with generality and robustness. In contrast, the precoding matrix should be designed according to the currently optimized antenna positions. Therefore, $f_{\text{BF}}(\cdot)$ should be a function of both the user feedback bits and the optimized MAs' positions at a specific time slot.


It is worth noting that problem \((\mathcal{P}2)\) exhibits a structure similar to that of \((\mathcal{P}1)\), and thus the solution methodology outlined in Section \ref{sec:Proposed mechanism} can be directly applied. The main distinction is that the MAs' positions must be optimized over a longer time period, i.e., based on a set of CSI, and then held fixed during that interval, whereas \(f_{\text{BF}}(\cdot)\) is updated on an instantaneous basis using current feedback. 

\subsection{MAs' Positions Optimization under Statistical CSI}\label{Stat-DNN}

Under the assumption of statistical CSI, the optimization of MA positions for a given \(\mathbf{Q}^{\rm all}\) reduces to learning a fixed antenna configuration that is optimal for the current channel distribution. Specifically, the antenna position optimization module is designed to capture the underlying features from multiple samples of user feedback bits and to output a deterministic MAs' positions.

To achieve this, we first introduce an encoder that takes as input a collection of user feedback data accumulated over \(T_{\delta}\) time slots, resulting in a tensor of size \(T_{\delta} \times K \times B\). The encoder extracts temporal and spatial correlations embedded in the feedback sequence, thereby learning representative features of the statistical channel environment. These features are then passed through a multi-layer DNN, which further refines the representation and produces a deterministic MAs' positions via a processing scheme similar to that described in Section~\ref{subsec: DNN Rep}. The overall processing pipeline can be summarized as  
\begin{equation}
\mathbf{p}_{\mathcal{G}}^s = \mathcal{Z}\bigl(\mathcal{T}(\mathbf{Q}^{\rm all})\bigr),
\end{equation}  
where \(\mathcal{T}(\cdot)\) denotes the feature extraction encoder, and \(\mathcal{Z}(\cdot)\) is a fully connected DNN. The activation function at the output layer of \(\mathcal{Z}\) is a temperature-controlled softmax function, identical to the one defined in \eqref{T-softmax}. The hard selection vector $\mathbf{m}^s$ can be constructed by setting the entries corresponding to the top-$N$ elements of $\mathbf{p}_{\mathcal{G}}^s$ to one and the rest to zero.
Similarly, to obtain a hard antenna selection result, we apply the following operation
\begin{equation}
\mathbf{m}^s = \mathbf{p}_{\mathcal{G}}^s + \mathcal{D}\left( \mathbf{m}^s - \mathbf{p}_{\mathcal{G}}^s \right),
\end{equation}  
where \(\mathcal{D}(\cdot)\) denotes the ST estimator, which ensures differentiability during backpropagation while enabling hard selection during the forward pass.

\begin{figure}[!t]
  \centering
  \includegraphics[width=0.96\linewidth]{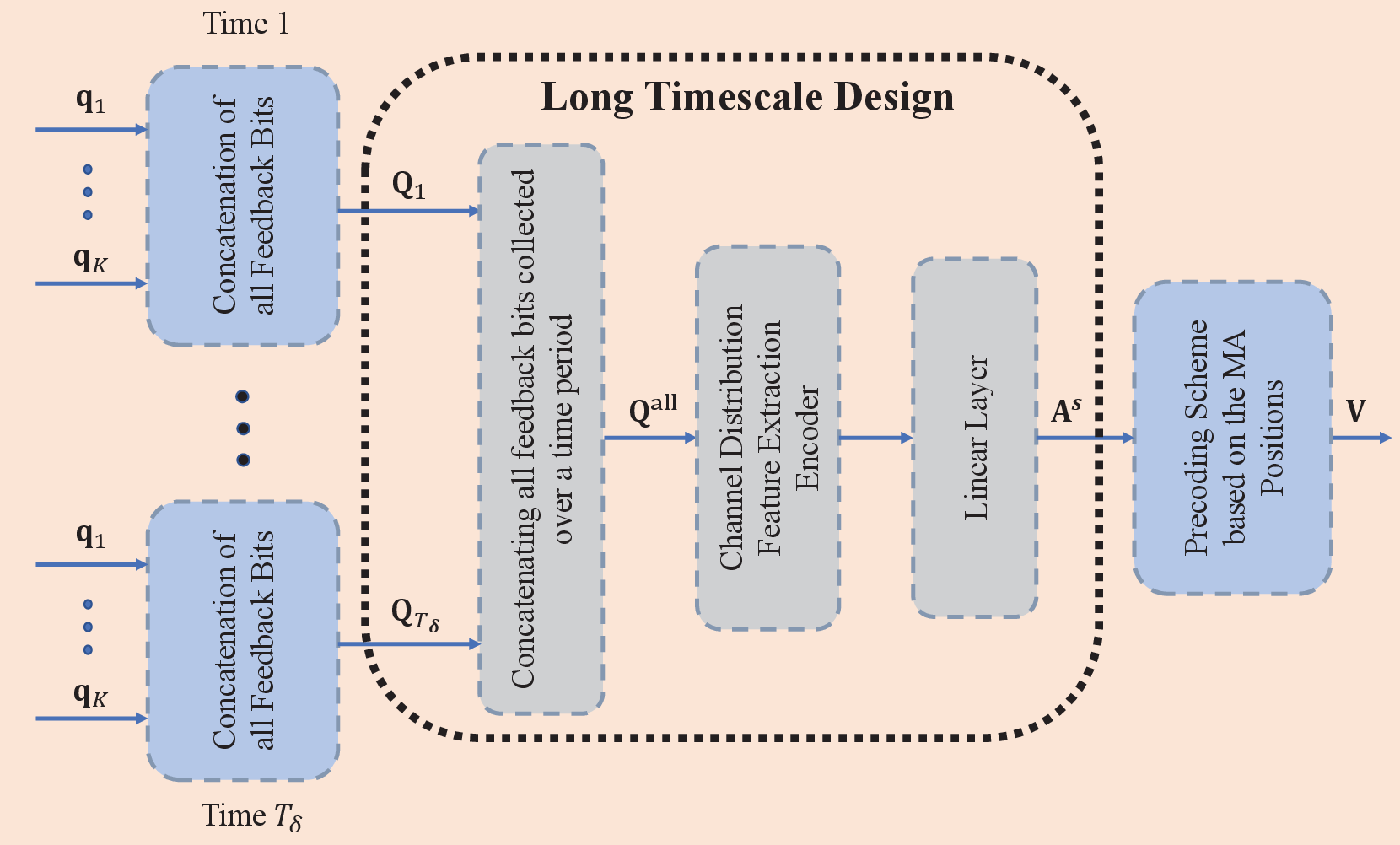}
  \caption{Illustration of the MA position optimization module under the statistical channel model.}
  \label{fig:stat_MA_opt}
  \vspace{-1.75em}
\end{figure}



\section{Implementation Details}\label{sec:Implementation Details}

We implement the proposed framework using Torch\cite{torch} and follow the training procedure in Section \ref{sec:Proposed mechanism} and Section \ref{Stat-DNN} to learn the parameters of the combined DNNs. In this section, we provide the implementation details of the proposed DNN in Fig. 2 and its training procedure in Algorithm \ref{alg:joint_ma_dnn}.

We employ a SGD-based training algorithm, namely the Adam optimizer\cite{adam}, with a mini-batch size of $|\mathcal{B}| = 32$ and a learning rate $\eta$ that progressively decays from $10^{-4}$ to $10^{-6}$. On the user side, a 3-layer fully-connected deep neural network (DNN) is used to compress the received pilot signal into a sequence of feedback bits. The numbers of hidden neurons in the three layers are $[l_1, l_2, l_3] = [1024, 512, B]$. To accelerate convergence and improve training stability, each dense layer is followed by a batch normalization layer. On the BS side, a unified CNN-based architecture is adopted to jointly infer the MAs' positions and precoding matrix from users feedback bits: the input is a $B \times K$-dimensional feedback bit matrix collected from $K$ users, which is reshaped into a 1-D format and passed through three sequential convolutional layers. Each convolution uses a kernel size of 3 with padding of 1, and is followed by batch normalization. The first convolutional layer outputs 1024 channels, the second increases the dimensionality to 2048 channels, and the third reduces it back to 1024 channels. The output features are then flattened and fed into a fully-connected layer with 1024 neurons, again followed by batch normalization. 
For the MA positions optimization DNN illustrated in Fig.~\ref{fig:system}, we first employ a feature extraction module composed of a  Transformer encoder \cite{transformer} and a learnable CLS token \cite{clstoken}, which captures temporal CSI features from the feedback sequence. Subsequently, a fully connected layer with 512 neurons is used to project the extracted features into a $G$-dimensional vector.

In order to optimize the downlink training pilot matrix $\tilde{\bf X}$, we define \(\mathbf{X}\) as a training variable in Torch whose initial value is randomly generated according to an i.i.d. complex Gaussian distribution with zero mean and variance \(\sqrt{P_{\text{max}}/N}\), such that the transmitted pilots in the \(l\)-th pilot transmission satisfy the power constraint. To ensure the block-diagonal structure of the pilot $\tilde{\mathbf{X}}$ as shown in \eqref{diagX}, we introduce a matrix with form
\begin{equation}
\mathbf{M} = 
\begin{bmatrix}
\mathbf{1}_{N\times Z} & \mathbf{0} & \cdots & \mathbf{0} \\
\mathbf{0} & \mathbf{1}_{N\times Z} & \cdots & \mathbf{0} \\
\vdots  & \vdots  & \ddots & \vdots  \\
\mathbf{0} & \mathbf{0} & \cdots & \mathbf{1}_{N\times Z}
\end{bmatrix}
\end{equation}
where $\mathbf{1}_{N \times Z}$ denotes an all-ones matrix of size $N \times Z$. The pilot matrix $\textbf{X}$ is then multiplied by $\mathbf{M}$ to impose the desired block-diagonal structure. Finally, $\textbf{X}$ is normalized to satisfy the transmit power constraint.

To mitigate the risk of local optima arising from the joint optimization of the MAs' positions and the precoding matrix, we adopt an alternating training strategy: we first fix the parameters \(\Theta_A\) of the MA position optimization network, and optimize the remaining parameters \(\{\textbf{X}, \{\Theta_R^{(k)}\}_{k=1}^K, \Theta_T\}\). In the subsequent phase, we allow \(\Theta_A\) to be learnable and jointly optimize all network parameters. This two-stage training enables the precoding network to adapt to the current antenna configuration, thereby avoiding situations where the precoding scheme is not aligned with the MA positioning mechanism. Moreover, it helps prevent the precoder from overfitting or getting trapped in suboptimal solutions tied to specific antenna configurations. The two training modes are alternated every two epochs during the training process.


To generate the training dataset, we consider a cellular environment spanning a three-dimensional region defined by ${X,Y,Z} = [0, 100] \times [0, 100] \times [-5, 5] $ meters. A BS is positioned at the fixed location of $[50, 0, 10]$ meters. Within this environment, a large number of scatterers are randomly distributed to emulate a realistic multipath propagation scenario. The transmit FRM at the BS, as well as the PRM between the BS and the users, are assumed to remain invariant over time. A total of $K$ mobile users move randomly within the coverage area. At each time step, the channel is simulated by randomly assigning each user to a subset of nearby scatterers, and calculating the corresponding receive-side FRMs based on the geometric relationships. 
We adopt a two-component Rician model (BS-user direct path as LoS; scatterer paths as NLoS). Let the unscaled expected powers be $\bar P_{\mathrm{LoS}}$ and $\bar P_{\mathrm{NLoS}}$ with $\bar P_{\mathrm{total}}=\bar P_{\mathrm{LoS}}+\bar P_{\mathrm{NLoS}}$. For a Rician factor $\beta$, we scale the two parts by $\eta_{\mathrm{LoS}}=\big(\tfrac{\bar P_{\mathrm{total}}}{\bar P_{\mathrm{LoS}}}\tfrac{\beta}{1+\beta}\big)^{1/2}$ and $\eta_{\mathrm{NLoS}}=\big(\tfrac{\bar P_{\mathrm{total}}}{\bar P_{\mathrm{NLoS}}}\tfrac{1}{1+\beta}\big)^{1/2}$, guarantees $\eta_{\mathrm{LoS}}^{2}\bar P_{\mathrm{LoS}}/(\eta_{\mathrm{NLoS}}^{2}\bar P_{\mathrm{NLoS}})=\beta$ while preserving the total expected power $\eta_{\mathrm{LoS}}^{2}\bar P_{\mathrm{LoS}}+\eta_{\mathrm{NLoS}}^{2}\bar P_{\mathrm{NLoS}}=\bar P_{\mathrm{total}}$.

To construct the training data, we uniformly sample the movable region with a spacing of $\lambda/2$ to obtain $M$ candidate antenna positions for channel measurements. The corresponding complex-valued dataset $\textbf{h} \in \mathbb{C}^{M \times K}$, representing the channels between these candidate antennas and $K$ users, is used as the input to the DNN for channel estimation. In addition, we sample the movable region with a finer resolution of $\lambda/4$ to obtain a denser set of channel responses, forming the ground-truth channel matrix $\textbf{H} \in \mathbb{C}^{G \times K}$, which is used for computing the loss function defined in~\eqref{eq:loss function}. The learning objective is to minimize this loss on the training dataset with 10\% of them is used to validate. A total of 30{,}000 training samples are generated based on this specified channel distribution to train the proposed network, with 10\% of them reserved for validation. 
The system parameters are configured as follows: pilot length per round is set to \(L = 8\), the Rician factor is $\beta = 10$dB, and the maximum transmission power is \(P_{\max} = 1\)~W. The noise power satisfies $10\log_{10}(\frac{P_{\rm max}}{\sigma^2}) = 20$dB. Regarding the training setup, we adopt a total of \(N_{\rm ep,max} = 1000\) training epochs, the initial annealing rate and learning rate are set to \(\omega^{(0)} = 1\) and \(\eta^{(0)} = 0.0001\), respectively. The annealing rate is updated at each epoch according to \(\omega^{(i)} = \min(1.01 \times \omega^{(i-1)}, 10)\), while the learning rate is adjusted as \(\eta^{(i)} = \max(0.99 \times \eta^{(i-1)}, 10^{-6})\).

\section{Simulation Results}\label{sec:Simulation}

We now evaluate the performance of the proposed framework for MA-aided downlink systems. The proposed method is compared with the following baselines (BL):

\textit{1) DL-based joint design of MAs' positions and precoders:} A DNN architecture, structurally similar to the proposed framework but without the pilot design and quantization modules, is trained using identical training parameters. The DNN jointly outputs both the MAs' positions and corresponding precoders. This method is evaluated under both perfect CSI and estimated CSI scenarios. In the estimated CSI case, the channel estimation procedure is conducted as follows: MMSE-based channel estimation is performed at each pre-defined candidate position, resulting in an estimated channel matrix \(\mathbf{H} \in \mathbb{C}^{M \times K}\), which serves as the input to the proposed DNN. To comply with hardware constraints, the downlink channel estimation is carried out over \(Z\) training rounds, where the BS transmits a pilot matrix \(\hat{\mathbf{X}} \in \mathbb{C}^{N \times L}\) in each round. The pilot matrix is chosen as a DFT matrix~\cite{MMSE-MIMO}.

\textit{2) Gradient-based MAs' positions method}  This benchmark adopts the gradient descent method proposed in~\cite{ref22} to optimize the MAs' positions, based on the estimated CSI. The corresponding precoding matrix is computed using the conventional
zero-forcing (ZF) algorithm~\cite{ZF-classic}. This method can operate with either perfect CSI or estimated CSI to optimize MAs' positions over the continuous region.

\textit{3) Antenna selection (AS) and ZF (AS-ZF):} The continuous movable region is discretized into a grid with spacing \(\lambda/2\), and an efficient antenna selection algorithm proposed in~\cite{AntennaSelection} is applied to obtain a suboptimal set of MA positions. The precoder is then computed using the ZF method. This benchmark is evaluated using estimated CSI to be aligned with the conventional methodologies.

\textit{4) Fixed antenna position and ZF (Fixed-ZF) :} MAs' positions are fixed to form a uniform linear array. ZF is applied to obtain the precoding matrix. Similar to \textit{1)}, this method is evaluated using both perfect and estimated CSI.

\begin{figure}[t!]
	\centering
	\begin{subfigure}[b]{0.239\textwidth}
		\centering
		\includegraphics[width=\textwidth,height=0.73\textwidth]{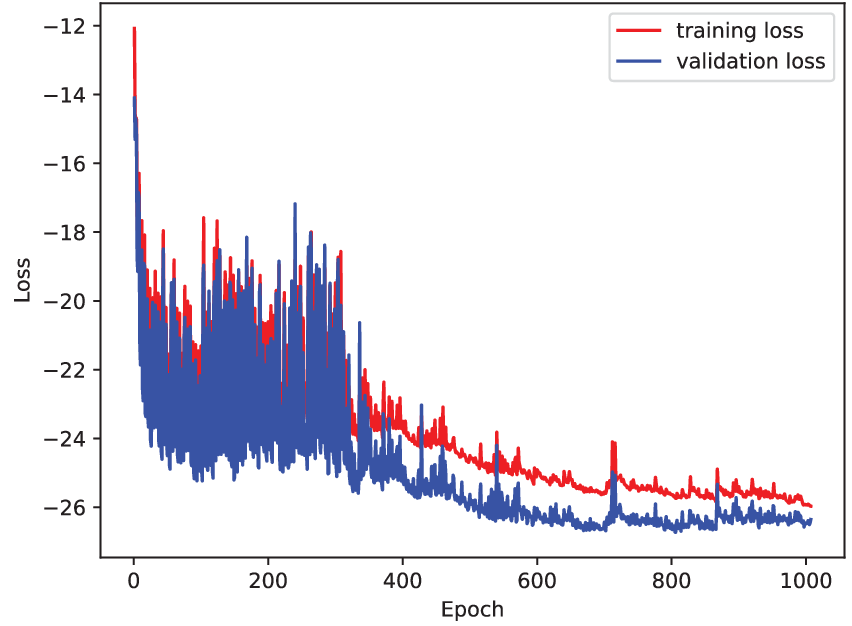}
		\caption{}
		\label{fig:training_3lambda}
	\end{subfigure}
	\hfill
	\begin{subfigure}[b]{0.239\textwidth}
		\centering
		\includegraphics[width=\textwidth,height=0.73\textwidth]{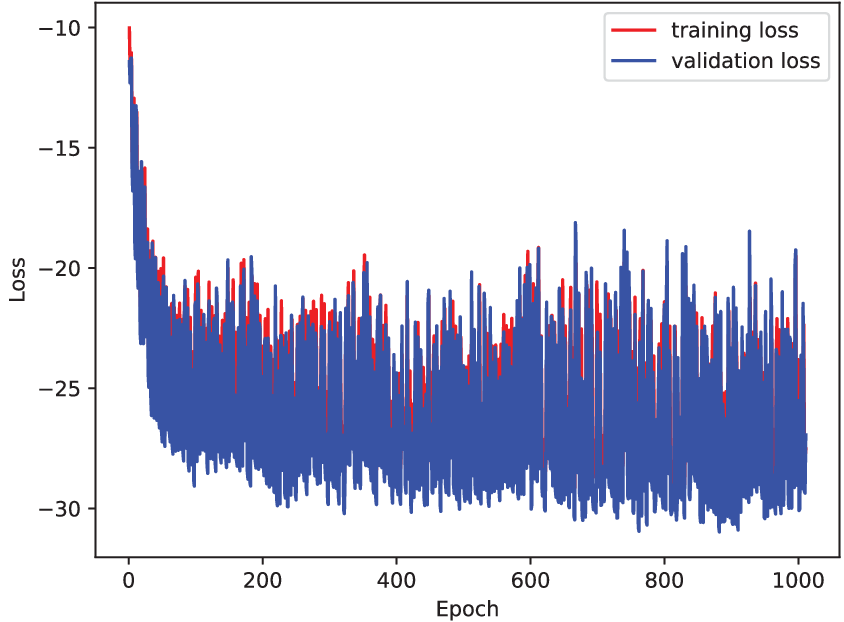}
		\caption{}
		\label{fig:training_5lambda}
	\end{subfigure}
	\caption{Training convergence with \(K = 8\) in a movable antenna region of size (a) \(3\lambda \times 3\lambda\), (b) \(5\lambda \times 5\lambda\).}
	\label{fig:training_comparison}
	\vspace{-1.75em}
\end{figure}
Fig.~\ref{fig:training_3lambda} and Fig.~\ref{fig:training_5lambda} illustrate the  convergence behavior of the proposed DNN under different movable regions. The networks achieves a significant loss reduction over epochs, indicating successful joint learning of MAs' positions optimization and precoding. When the movable region is enlarged from \(3\lambda \times 3\lambda\) to \(5\lambda \times 5\lambda\), the final loss becomes notably lower, which highlights the benefit of a larger feasible domain for MA placement. The expanded search space allows the network to explore more favorable MA configurations, thus improving overall performance. Nevertheless, the convergence in the \(5\lambda \times 5\lambda\) setup exhibits increased oscillation, which can be attributed to the stronger coupling between MAs' positions and precoders in a larger domain, as well as the heightened complexity of optimization due to a wider solution space. Despite the fluctuations, the overall coverage confirms the robustness and adaptability of the proposed training framework.

\begin{figure}[t!]
\centering 
 \includegraphics[width=3.36in]{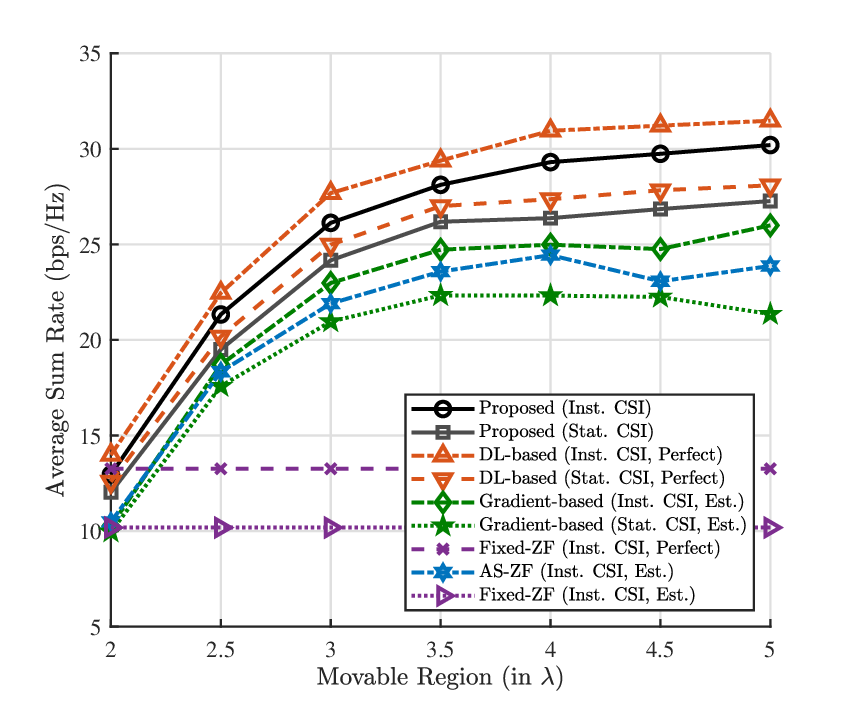} 
\caption{Comparison of system performance under different movable region sizes, with \(N = 16\).}
   \label{fig:inst_G_16}
   \vspace{-1.5em}
\end{figure}
\begin{figure}[t!]
\centering 
 \includegraphics[width=3.36in]{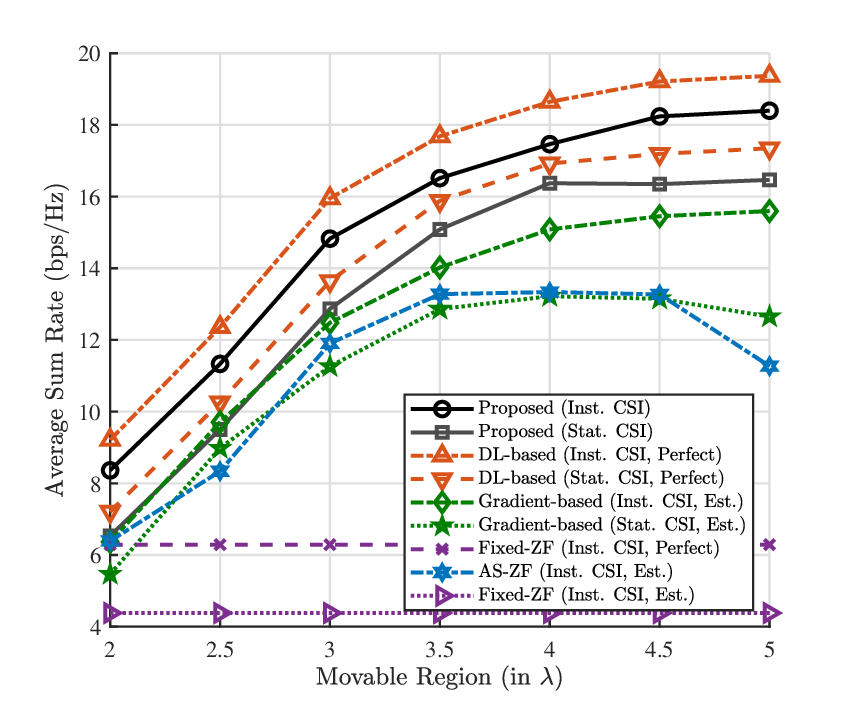} 
\caption{Comparison of system performance under different movable region sizes, with \(N = 8\).}
   \label{fig:inst_G_8}
   \vspace{-1.75em}
\end{figure}
Fig.~\ref{fig:inst_G_16} and Fig.~\ref{fig:inst_G_8} present the average sum rate performance of the proposed DNN-based framework and various BLs for a system in which $K = 8$, $L_{k}^t = L_{k}^r = 6$ under different movable region sizes, for \(N = 16\) and \(N = 8\) MAs, respectively. 
Under the statistical CSI, the proposed framework achieves performance that closely approaches the DL-based method under perfect CSI, which serves as an upper bound. Furthermore, compared to its performance under the instantaneous CSI scenario, the proposed scheme exhibits only a slight degradation, demonstrating strong robustness and adaptability in statistical channel environments. 
Specificly, compared with AS-ZF, which relies on static heuristics and outdated feedback, whereas our offline-trained DNN mapping yields near-optimal MA positions and achieves higher sum-rates.
Besides, it can be observed that the proposed framework consistently outperforms all BLs which involve downlink training for channel estimation, including those based on MMSE estimation and traditional antenna selection. Although its performance is slightly inferior to that of the ideal benchmarks assuming perfect CSI with joint optimization of MAs' positions and precoding 
the gap of performance is acceptable because that the difficulty of acquiring accurate CSI in MA systems. This challenge becomes more pronounced as the size of the movable region increases, which incurs additional channel estimation overhead and performance degradation. 
In addition, BLs relying on estimated channel exhibit diminishing returns or even degradation in performance as the movable region grows, due to the increased sensitivity to estimation errors and pilot overhead. In contrast, the proposed DNN demonstrates strong robustness, effectively learning to compensate for estimation inaccuracies through data-driven training. 

\begin{figure}[t!]
\centering 
 \includegraphics[width=3.36in]{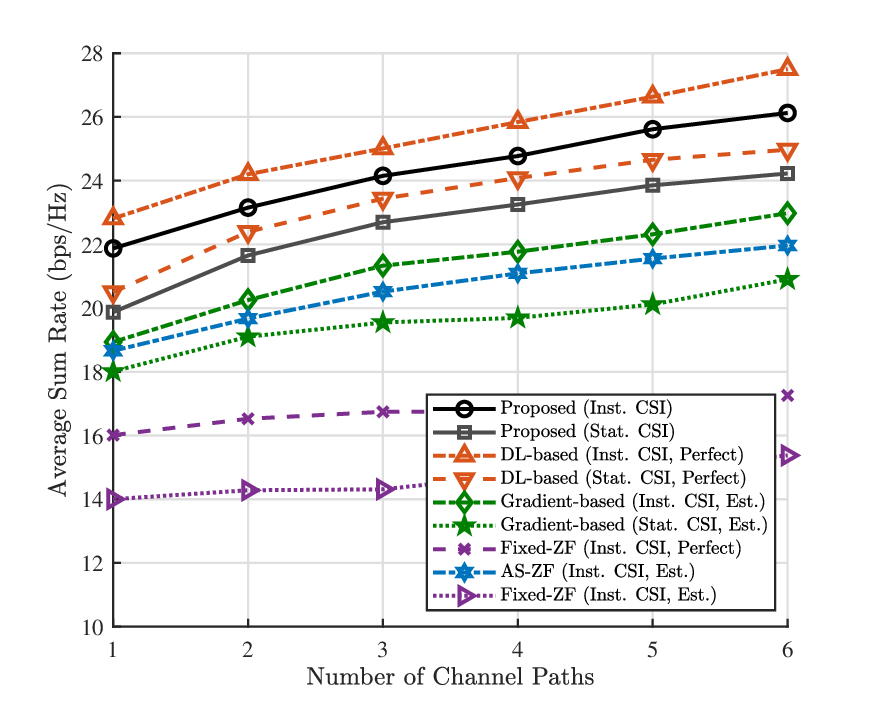} 
\caption{Performance comparison versus number of channel paths.}
   \label{fig:inst_L}
   \vspace{-2em}
\end{figure}
Fig.~\ref{fig:inst_L} illustrates the system performance versus the number of channel paths, assuming an equal number of transmit and receive paths, i.e., \(L_k^t = L_k^r\). As the number of paths increases, all methods experience performance gains due to increased spatial diversity~\cite{ref4}.
The proposed method maintains a strong performance advantage over all benchmarks relying on estimated channel. While DL-based method under perfect CSI still achieve the highest sum rate, the proposed DNN-based solution closely tracks its performance. Since the training data include samples with different numbers of channel paths, the network learns to generalize across channel environments without requiring retraining for each specific setting. This property significantly enhances its practical applicability, particularly in mobile or heterogeneous environments where the channel sparsity level may vary over time. Hence, the proposed mechanism not only performs well under static conditions but also adapts effectively to dynamic propagation environments characterized by variable path richness. For instance, when the number of channel paths is as high as \(L_k^t = L_k^r = 6\), the proposed method under instantaneous CSI achieves approximately 95.86\% of the average sum rate of the upper bound (Inst. CSI, Perfect), while the version based on statistical CSI reaches around 96.80\% of the upper bound (Stat. CSI, Perfect). This indicate that under sparse multipath conditions, the proposed method can effectively exploit limited channel diversity and maintain performance close to the theoretical upper bounds.
 
\begin{figure}[t!]
\centering 
 \includegraphics[width=3.36in]{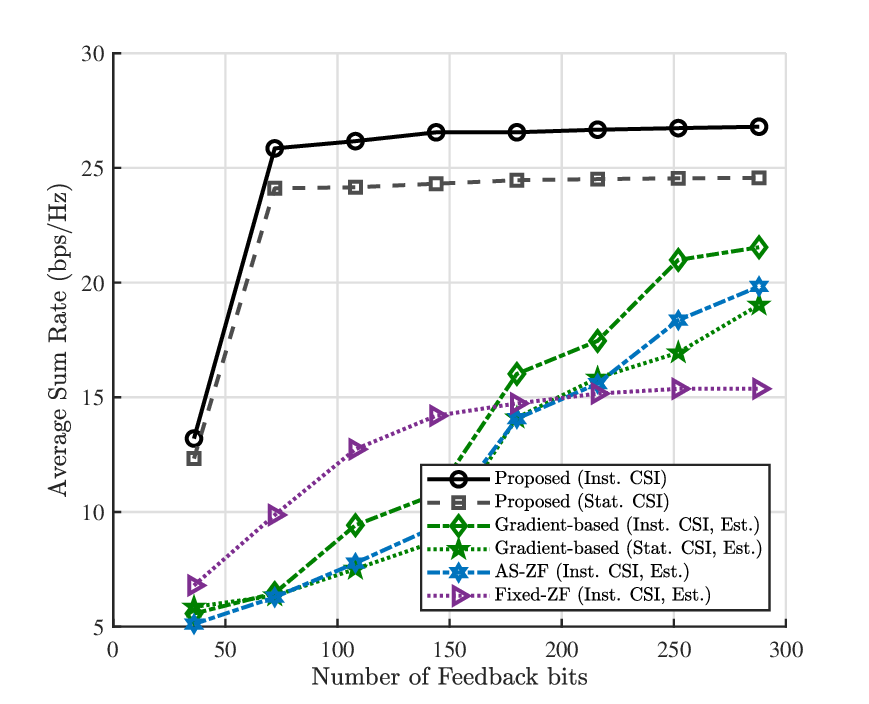}
\caption{Performance comparison versus number of feedback bits.}
\vspace{-1.75em}
   \label{fig:inst_B}
\end{figure}
Fig.~\ref{fig:inst_B} evaluates the impact of feedback bit number on the average sum rate, highlighting the effectiveness of the proposed quantization and feedback strategy. To simulate practical scenarios, all BLs that rely on downlink channel estimation are extended with a quantization stage: after estimating the channel at the user, the complex-valued of CSI is quantized into a finite bit representation, transmitted as a bitstream, and then dequantized at the BS to obtain the CSI. This setup allows us to quantify the performance degradation caused by limited feedback.
As observed in Fig.~\ref{fig:inst_B}, the proposed DNN-based mechanism significantly outperforms all BLs across different feedback bit levels, particularly in the low-bit regime. By embedding a learned quantization mechanism within the network, the proposed framework efficiently compresses the estimated CSI into compact bitstreams while preserving essential channel information. Unlike conventional schemes that rely on hand-crafted quantization followed by bit-to-channel recovery, the proposed method jointly optimizes the encoder-decoder process, effectively learning a task-oriented and efficient representation. Moreover, the proposed method achieves near-saturation performance with substantially fewer feedback bits compared to traditional methods, indicating its superior bit efficiency and scalability in feedback-limited systems. 

\begin{figure}[t!]
\centering 
 \includegraphics[width=3.36in]{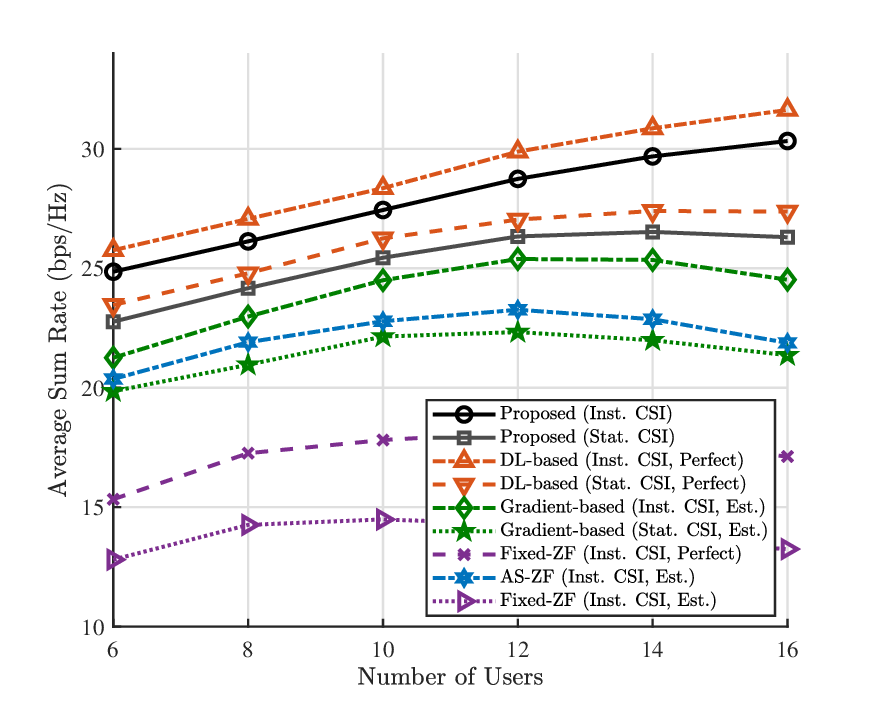} 
\caption{Performance comparison versus number of Users.}
\vspace{-1.5em}
   \label{fig:inst_K}
\end{figure}
Fig.~\ref{fig:inst_K} shows the average sum rate performance of different schemes as the number of users increases. The proposed method continues to exhibit strong performance under both instantaneous and statistical CSI models, outperforming all BLs that rely on estimated CSI.
As the number of users increases, the performance gap between the proposed framework and the BLs becomes more pronounced. In particular, schemes designed under the statistical CSI tend to experience either a slowdown in performance growth or an actual decline. This can be attributed to two primary factors. First, statistical CSI-based designs have limited ability to accurately capture and mitigate inter-user interference as number of user increases. Second, for schemes which rely on estimated CSI, the accumulation of estimation errors across users further exacerbates performance degradation, such as AS-ZF and Fixed-ZF.
For example, when the number of users increases from \(K = 14\) to \(K = 16\), the performance of gradient-based method (Stat. CSI, Est.) and AS-ZF (Inst. CSI, Est.) shows a small drop, highlighting their sensitivity to channel estimation errors and limited interference modeling capability. In contrast, the proposed mechanism under the statistical CSI model (Proposed, Stat. CSI) maintains a steady upward trend, with only a marginal performance gap relative to the instantaneous CSI counterpart. This demonstrates its superior robustness and adaptability in multiuser environments, even without relying on instantaneous channel knowledge.

%

\section{Conclusion}\label{concl}
This paper proposes a DL-based framework for the joint design of channel estimation, antenna position optimization, and precoding in MA-aided downlink communication systems. Specifically, an end-to-end trainable neural network is developed, which integrates key phases of the physical layer, including downlink pilot design, user-side channel estimation and quantization-based feedback, as well as joint MA position optimization and precoding at the BS. Unlike conventional designs that separately optimize each module under idealized assumptions, the proposed architecture learns to coordinate all stages in a data-driven manner. Furthermore, transformer model is introduced as a temporal channel feature extractor, extending the proposed framework to a two-timescale design and thereby reducing the antenna movement overhead.
Comprehensive simulation results demonstrate that the proposed method achieves superior performance compared to traditional baseline schemes under both instantaneous and statistical channels. 
Future work may explore the extension of the framework to multi-cell scenarios, real-time deployment under hardware constraints, and the incorporation of sensing functionalities into the joint design.

\end{document}